\documentclass[11pt,secnumarabic,aps,amsmath,amssymb,amsfonts,superscriptaddress,tightenlines,nobibnotes,prd,nofootinbib,twocolumn,dvipsnames]{revtex4}

\usepackage{bm}
\usepackage[utf8]{inputenc}
\usepackage{amsmath}
\usepackage{amssymb}
\usepackage{appendix}
\usepackage{pgfplots}
\usepackage{wasysym}
\usepackage{subeqnarray}
\usepackage{multirow}
\usepackage{graphicx}
\usepackage{xcolor}
\usepackage{epsfig}
\usepackage[hidelinks,colorlinks=true,citecolor=Blue,linkcolor=Blue,urlcolor=BlueViolet]{hyperref}

\numberwithin{equation}{section}

\usepackage{cancel}
\usepackage[normalem]{ ulem }
\usepackage{soul}

\newcommand\spart{\;\raise1.0pt\hbox{/}\hskip-6pt\partial}
\newcommand\spartb{\;\overline{\raise1.0pt\hbox{/}\hskip-6pt\partial}}

\newcommand{\be}{\begin{equation}}
\newcommand{\ee}{\end{equation}}
\newcommand{\bea}{\begin{eqnarray}}
\newcommand{\eea}{\end{eqnarray}}
\newcommand{\beal}{\begin{align}}
\newcommand{\eeal}{\end{align}}
\newcommand{\beas}{\begin{subeqnarray}}
\newcommand{\eeas}{\end{subeqnarray}}
\renewcommand\[{\begin{equation}}
\renewcommand\]{\end{equation}}
\newcommand{\dd}{{\rm d}}
\newcommand{\ii}{{\rm i}}

\newcommand{\HH}{{\cal H}}

\newcommand{\gr}[1]{\boldsymbol{#1}}
\newcommand{\ellc}{\ell_c}
\newcommand{\ells}{\ell_s}
\newcommand{\csch}{{\rm csch}}
\newcommand{\I}{{{\rm I}}}
\newcommand{\V}{{{\rm V}}}
\newcommand{\VIIo}{{{\rm VII}_0}}
\newcommand{\VIIh}{{{\rm VII}_h}}
\newcommand{\IX}{{{\rm IX}}}
\newcommand{\Rspatial}{{{}^{(3)}\! R}}
\newcommand{\hatD}{{\cal D}}

\newcommand{\barK}{{\cal K}}
\newcommand{\myzeta}{{\zeta}}
\newcommand{\mygamma}{{\xi}}

\newcommand{\com}[1]{{}}

\newcommand\cpsilent{}

\begin{document}

\title{Bianchi spacetimes as super-curvature modes around isotropic cosmologies}

\author{Thiago S. Pereira}
\email[]{tspereira@uel.br}
\affiliation{Departamento de Física, Universidade Estadual de Londrina, Rod. Celso Garcia Cid, Km 380, 86057-970, Londrina, Paraná, Brazil.}
\author{Cyril Pitrou}
\email[]{pitrou@iap.fr}
\affiliation{Institut d'Astrophysique de Paris, CNRS UMR 7095, 98 bis Bd Arago, 75014 Paris, France.}
\date{\today}
\begin{abstract}
A powerful result in theoretical cosmology states that a subset of anisotropic Bianchi models can be
seen as the homogeneous limit of (standard) linear cosmological perturbations. Such models are precisely those leading to Friedmann spacetimes in the limit of zero anisotropy. Building on previous works, we give a comprehensive exposition of this result, and perform the detailed
identification between anisotropic degrees of freedom and their corresponding scalar, vector, and tensor perturbations of standard
perturbation theory. In particular, we find that anisotropic models very close to open (i.e., negatively curved) Friedmann spaces correspond to some type of super-curvature perturbations. As a consequence, provided anisotropy is mild, its effects on all types of cosmological observables can
always be computed as simple extensions of the standard techniques used in relativistic perturbation theory around Friedmann models. This 
fact opens the possibility to consistently constrain, for all cosmological observables, the presence of large scale anisotropies on the top of the 
stochastic fluctuations.
\end{abstract}
\maketitle


\section{Introduction}\label{sec:Intro}

Homogeneous and spatially anisotropic cosmologies, commonly referred
to as Bianchi models, have long been the arena for 
new developments in theoretical cosmology. The interest in these models stem from their unique ability to preserve a high degree of symmetry while remaining 
phenomenologicaly versatile. Despite the fact that CMB data seems to favor the more restricted class of isotropic Friedmann-Lemaître-Robertson-Walker (FLRW) universes, Bianchi models are possibly the simplest extensions of a maximally symmetric expanding universe, and for that reason they are theoretically (if not observationaly) interesting.  However, despite all of their attractiveness, one cannot avoid the feeling that Bianchi models fall in the category of ``alternative cosmologies''.

Meanwhile, a robust but less known theoretical result states that all nearly isotropic Bianchi models with isotropic limit (namely, models $\I$, $\VIIo$, $\V$, $\VIIh$ and $\IX$) can be extracted from standard (i.e., FLRW) cosmological perturbations in the limit that these perturbations become homogeneous. This idea, explored in the case of model $\IX$ in Refs.~\cite{Grishchuk:1975ec,King:1991jd}, and fully established for the other models by Pontzen and Challinor in 2010 (henceforth PC10)~\cite{Pontzen2010}, bridges the gap between FLRW and Bianchi models, forcing us to see the latter as legitimate manifestations of the standard cosmological framework. 

The idea that Bianchi models emerge as homogeneous cosmological
perturbations on the top of an isotropic universe is quite
intuitive. In fact, since linear cosmological perturbations break both
translational and rotational isometries of the FLRW background metric,
a (suitably defined) homogeneous limit of these perturbations should
restore translational invariance while keeping the most general
spatial anisotropies compatible with homogeneity. The remaining
anisotropic degrees of freedom are, then, nothing else but those
describing the subset of Bianchi models --- exactly the subset having
the initial FLRW metric as their isotropic limits. In fact, by
properly defining the isometries of the FLRW metric and demanding
rotational invariance to be broken, one can build the isometries of
the corresponding Bianchi models from first
principles~\cite{Pontzen2010}. This idea not only leads to a more
intuitive formulation of Bianchi models~\footnote{However, it does not
apply to Bianchi models not having FLRW limits, namely, models ${\rm II}$,
${\rm III}$, ${\rm IV}$, ${\rm VI}_0$, ${\rm VI}_h$, and ${\rm VIII}$. It also excludes the homogeneous and anisotropic Kantowski-Sachs model, which falls outside the usual Bianchi classification anyway.} but, more importantly, sheds light on their connections with linear cosmological perturbations. Given the omnipresence of linear perturbation theory in the cosmologist's toolkit, the exploration of these connections becomes central for a better understanding of the $\Lambda$CDM model.

One simple example of this connection is easily illustrated: the spatial anisotropies of Bianchi type $\I$ model is dynamically equivalent to a gravitational wave of infinite wavelength (i.e., homogeneous) on the top of a spatially flat FLRW universe. Indeed, for small anisotropies, the Bianchi~$\I$ metric reads
\be
\dd s^2 = a^2(\eta)[-\dd\eta^2+(\delta_{ij}+2\beta_{ij}(\eta))\dd x^i\dd x^j]\,,\nonumber
\ee
whereas a FLRW universe with linear and homogeneous gravitational wave is described by
\be
\dd s^2 = a^2(\eta)[-\dd\eta^2+(\delta_{ij}+2E_{ij}(\eta))\dd x^i\dd x^j]\,.\nonumber
\ee
Since both the shear $\beta_{ij}$ and the wave amplitude $E_{ij}$ are symmetric and trace free, both evolve as
\be
X''_{ij}+2\frac{a'}{a}X'_{ij}=0\,.\nonumber
\ee

While the above analogy could at first look as a happy accident of model $\I$, it is actually not. The general correspondence in the weak field regime was investigated for the five Bianchi types with a FLRW limit in PC10. As hinted by these authors, and explicitly demonstrated here, the connection results in a richer structure than could have been expected from the simple example above, since some Bianchi models arise as \emph{finite} wavelength perturbations over FLRW spacetimes. Indeed, the homogeneous limit of models $\VIIo$, $\V$, $\VIIh$, and $\IX$ is not simply given (in Fourier space) by $k\rightarrow0$, but rather by a proper identification of an effective mode $\nu_m$ defined in terms of the Fourier mode $k$ as
\be
\nu_m^2 = k^2 + (1+|m|)K\,,\nonumber
\ee
where $|m|=0,1,2$ accounts respectively for scalar, vector and tensor
perturbations, and $K$ is the spatial curvature of the corresponding
FLRW model. Furthermore, we show that for models $\V$ and $\VIIh$ the
homogeneous limit corresponds to a complex $\nu_m$ or, equivalently,
to a perturbation whose wavelength is larger than the curvature
scale. In this case, the construction of a proper eigenbasis for the
perturbations requires analytical continuation of radial
functions. The implementation and observational consequences of these
modes with this method will be explored in a forthcoming publication \cite{DualityLetter}.

The interplay between Bianchi and perturbed FLRW models also has important practical applications: if one wants to derive the dynamical behavior of some cosmological observable in a Bianchi~$\I$ universe, it is enough to derive the same dynamics for tensor perturbations in flat FLRW universe and take their homogeneous limit. This program was in fact used in~\cite{Adamek:2015mna} to derive theoretical expressions for the weak-lensing signal, and in~\cite{Marcori:2018cwn} to derive the 
direction and redshift time drifts of non-inertial observers. 

Here, we continue the effort started in PC10 and show that the same program can be applied to all Bianchi models with isotropic limit. In particular, we 
show how this Bianchi/Perturbed-FLRW duality can be used to infer predictions for any observable (as, e.g., CMB radiative transfer, weak gravitational lensing, etc.)
in Bianchi models from the well known methods of linearly perturbed
FLRW spacetimes~\cite{Kodama:1985bj,Mukhanov:1990me}. Hence, one can use the same theoretical framework by just separating modes describing global anisotropies from the ones describing stochastic perturbations. 

From the observational point of view, upper limits on the large scale anisotropy were placed in~\cite{Barrow:1985tda}, followed by claims of a Bianchi $\VIIh$
pattern in WMAP data \cite{Jaffe:2006sq,Jaffe:2005pw,McEwen:2013cka}. Further investigations using the method of~\cite{Pontzen2007,Pontzen2009} for radiative transport (see also the related method of \cite{Sung:2010ek}) combined with Planck data confirmed, however, that we can only obtain upper limits~\cite{Saadeh:2016bmp,Saadeh:2016sak} on the level of global anisotropy.

\begin{table*}
	\begin{ruledtabular}
		\begin{tabular}{clc}
			\textbf{Symbol} & \textbf{Definition} & Introduced in:\\
			\colrule
			$\{i,j,k,\cdots\}$ &  Spatial indices of non-coordinate (triad) basis & \S~\ref{subsec:general-construction} \\
			$\{a,b,c,\cdots\}$ &  Spacetime indices of non-coordinate (tetrad) basis & \S~\ref{subsec:general-construction} \\
			$\{\mu,\nu,\lambda,\cdots\}$ &  Abstract space-time indices& \S~\ref{subsec:general-construction} \\
			$\ii$ & Imaginary unity & \eqref{npm}\\
			$\gr{\xi}_i$ & Killing Vector Fields & \S~\ref{Sec2BianchiCosmologies}\\
			$\gr{e}_i$ & Invariant basis &  \S~\ref{Sec2BianchiCosmologies}\\
			$\gr{e}^i$ & Dual basis (or co-basis) to the invariant basis $\gr{e}_i$ &  \S~\ref{Sec2BianchiCosmologies}\\
			$e^\mu$ & Four-velocity of fundamental
                                  observers (used to foliate
                                  spacetime) &  \S~\ref{Sec2BianchiCosmologies}\\
			$u^\mu$ & Timelike fluid four-velocity. & \S~\ref{SecTmunu}\\
			\footnotesize{\it{FLRW quantities}}\\
			\hline
			$\gr{g}^{\rm MS}$ & Metric of maximally symmetric  spaces & \eqref{dsspat} \\
			$K$ & Spatial curvature of maximally symmetric spaces & \eqref{Defrchi} and Table~\ref{TableType} \\
			${\cal K}$ & Dimensionless spatial curvature & \eqref{dimensionlessK} \\
			$\hatD_i$ & Spatial covariant derivative associated with the FLRW metric & \eqref{SVT-metric}\\
            $\Delta$ & Laplace-Beltrami operator associated with $\hatD_i$ & \eqref{DefDelta} \\
            $Q^{(jm)}_{I_j}$ & Tensor \cpsilent{plane-wave} harmonic in position space & \eqref{laplace-beltrami} \\
            ${}_sG^{(jm)}_{\ell}$ & Normal modes in the total angular momentum representation & \eqref{DefGsjm} \\
            ${}_sY^{m}_{\ell}$ & Spin-weighted spherical harmonics & \eqref{DefGsjm} \\
            ${}_s\alpha^{(jm)}_\ell$ & Radial functions in the total angular momentum representation & \eqref{DefGsjm} \\
            ${}^\ell Q^{(jm)}_{I_j}$ & Tensor harmonic in  \cpsilent{$\ell$-representation}     &\eqref{Qljm} \\
            $\nu_m$&Modes of the Bianchi-FLRW    matching&        \eqref{qQplus}  and Table \ref{TableMatching}\\
             $\myzeta_\ell^m$   & Pseudo plane-wave weights for the Bianchi-FLRW matching& \eqref{qQplus}  and Table \ref{TableMatching}\\
			\footnotesize{\it{Bianchi quantities}}\\
			\hline
			$\gr{g}$ & Metric of Bianchi spacetime & \eqref{Bianchi-general-metric} \\
			$\gr{h}$ & Spatial metric of Bianchi spacetime   &   \eqref{Defh} \\
           	$\gamma_{ij}$ & Conformal spatial metric of Bianchi spacetimes  &       \eqref{Defgammaij} \\
			$\beta_{ij}$ & Expansion parameters of spatial  anisotropy & \eqref{def-betaij} \\
			${D}_i$ & Spatial covariant derivative in  Bianchi spaces & \eqref{DefD} \\            
            $q^{(m)}_{ij}$ & Polarization basis for shear $svt$ modes  &  \eqref{Defqij} \\
                       $N^{ij},A_i$ & Irreducible components  of the constants of structure&\eqref{DefNA}      and~\eqref{DefHatNHatA}
		\end{tabular}
	\end{ruledtabular}
	\caption{\label{list-of-symbols}List of main symbols used in
          this work.}
\end{table*}

We start this article by recalling some basic definitions of Bianchi
spacetimes in Section~\ref{Sec2BianchiCosmologies}, where we focus on the subset of models having 
isotropic limit. We then review, in Section~\ref{Sec4-SVT-FLRW}, some key
elements of linear perturbation theory in synchronous gauge, focusing
on the introduction of Scalar, Vector and Tensor modes and their
decomposition in terms of a complete basis of tensor harmonics. This
section summarizes the definition  and constructions of the companion paper~\cite{Pitrou:2019ifq}.
Moving forward, we introduce a set of linear modes for small anisotropies of Bianchi spacetimes in
Section~\ref{sec:bianchi-modes}; they are then used to find the exact
Bianchi/Perturbed-FLRW correspondence in
Section~\ref{SecMatch}. Finally, we discuss the cosmological implications of our
results in Section~\ref{Sec-Implications}.

Throughout this paper we use metric signature $(-,+,+,+)$ and units where $c=1$. Coordinate and 
non-coordinate indices, as well as a list of the main symbols encountered in this work, are defined
in Table~\ref{list-of-symbols}.

\section{Bianchi Cosmologies}\label{Sec2BianchiCosmologies}

Let us start with a brief and informal recap of spatially homogeneous (i.e., Bianchi) spacetimes. Detailed and 
pedagogical introductions can be found in many nice textbooks such as~\cite{ellis2012relativistic}, \cite{plebanski2006introduction} and~\cite{oyvind2007einstein}.

\subsection{General Construction}\label{subsec:general-construction}

Informally, a three-dimensional space is said to be homogeneous if for any pair of points there exists an isometric (i.e., metric-preserving) 
path connecting these points in a continuous way. The fields $\gr{\xi}$  tangent to such paths are Killing Vector Fields (KVFs), and are defined as
\be\label{DefKVF}
{\cal L}_{\gr{\xi}} \gr{g} = 0\,\quad \Leftrightarrow\quad \nabla_{(\mu} \xi_{\nu)}= 0\,,
\ee
where ${\cal L}$ is the Lie derivative and $\nabla$ is the covariant
derivative compatible with the metric $\gr{g}$. In three-dimensions,
the maximum number of such paths is 6, corresponding to 3 translations
and 3 rotations. However, because the space is three-dimensional,
there can be at most three KVFs which are \emph{everywhere} linearly independent. Any additional vector
$\gr{\zeta}$ obeying~\eqref{DefKVF} is necessarily of the form
$\gr{\zeta}=\sum_{i=1}^3 c_i(x)\gr{\xi}_i$. Given a point $p$ one can
form $\tilde{\gr{\zeta}} =\sum_{i=1}^3 [c_i(x)-c_i(p)]\gr{\xi}_i $ such that $\tilde{\gr{\zeta}}(p)=0$, which corresponds to a rotation around
point $p$. As an example, in the Euclidean three-dimensional space, $\gr{\zeta}=x^1\gr{\xi}_2-x^2\gr{\xi}_1$ corresponds to a rotation around the $x^3$-axis 
which keeps the point $p=(0,0,0)$ fixed. Thus, homogeneous and spatially anisotropic spaces are represented by KVFs which are everywhere linearly independent. 
Moreover, since the commutator of any two vectors $\gr{\xi}_i$ is another KVF, they form a closed algebra given by
\be\label{DefC}
[\gr{\xi}_i,\gr{\xi}_j]\equiv{C^k}_{ij} \gr{\xi}_k\,,
\ee
where the coefficients ${C^k}_{ij}$ are called the constants of structure of the algebra.

We can now build homogeneous spacetimes by simply stacking up homogeneous spaces, each of which labeled by a continuous time coordinate $t$ and having an orthogonal 1-form
\be\label{Defnfromt}
\gr{\omega} = -\dd t\,,\qquad\omega_\mu = -\partial_\mu t\,.
\ee
By construction, the vector $\gr{e}$ dual to $\gr{\omega}$ is orthogonal to the KVFs
\be\label{Propertyspatial}
\gr{e} \cdot \gr{\xi}_i = e^\mu g_{\mu\nu} \xi_i^\nu = 0\,,
\ee
and is normalized such that
\be\label{Propertyunit}
\gr{e}\cdot \gr{e} = e^\mu g_{\mu\nu} e^\nu = -1\,.
\ee

The task of finding explicit Bianchi spacetimes now consists of finding all constants ${C^k}_{ij}$ which are 
inequivalent under linear combination of the $\gr{\xi}_i$. This task is simplified by noting that, since ${C^k}_{ij}$
is antisymmetric in its lower indexes, it can be written as $C^{k}_{ij}\equiv \hat \epsilon_{ijl}H^{lk}$, where 
$\hat \epsilon_{ijl}$ is the permutation symbol (such that $\hat \epsilon_{123}=1$) and $H^{lk}$ is a general $3\times 3$ matrix. Decomposing the latter in 
its symmetric ($\hat N^{lk}$) and antisymmetric  ($\hat \epsilon^{lkm}A_m$) parts, we find that~\footnote{Our sign convention for $A_i$ agrees with that of PC10, but differs by a minus sign with that of Ref.~\cite{Pontzen2007}.}
\be\label{DefNA}
{C^k}_{ij} = \hat \epsilon_{ijl} \hat N^{lk} - A_i \delta_j^k + A_j \delta_i^k\,.
\ee
From the Jacobi identity 
\be
\hat \epsilon^{ijk}[\gr{\xi}_i,[\gr{\xi}_j,\gr{\xi}_k]] = 0
\ee 
the decomposition (\ref{DefNA}) implies
\be\label{NAzero}
\hat N^{ij}A_j = 0\,.
\ee
By suitable linear transformations of the $\gr{\xi}_i$ we can
diagonalize the matrix $\hat N^{ij}$ so that 
${\hat N^{ij} = {\rm diag}(N^1,N^2,N^3)}$. From property (\ref{NAzero}) we then see that $\gr{A}$ is either null or an eigenvector 
of the matrix $\hat{\gr{N}}$, and so we can set $\gr{A}=(0,0,A)$~\footnote{Note that our choice differs from the conventional one:
${\gr{A} = (A,0,0)}$~\cite{Ellis1969,Collins1973a,Barrow:2003fc,Pontzen2007}. One easily recovers the standard results 
just replacing the indexes $(1,2,3)$ in our expressions by $(2,3,1)$.}. We are thus left with
\begin{align}\label{ExplicitCommuteXi}
\begin{split}
[\gr{\xi}_1,\gr{\xi}_2] & = + N^3\gr{\xi}_3\,, \\
[\gr{\xi}_1,\gr{\xi}_3] & = - N^2\gr{\xi}_2+A\gr{\xi}_1\,, \\
[\gr{\xi}_2,\gr{\xi}_3] & = + N^1\gr{\xi}_1+A\gr{\xi}_2\,.
\end{split}
\end{align}

We now note that by suitably rescaling the lengths of the KVFs, we can set the components $N^1$, $N^2$, $N^3$, and $A$ to either $0$, 
$1$ or $-1$ (see, e.g. chapter 10 of~\cite{plebanski2006introduction}). We will not adopt this approach here. Instead, 
since the KVFs have dimensions of inverse length, we will keep these constants with the appropriate dimensions 
to maintain Eq.~\eqref{DefC} dimensionally homogeneous. For reasons that will become clear later, the constants $A$ and $N^i$
are associated with curvature and spiral lengths, respectively, so that we introduce 
\be\label{ells-def}
A \equiv \ellc^{-1}\,,\qquad N^i \equiv \ells^{-1}\,,
\ee
as two free parameters, except in the Bianchi $\IX$ case where $A=0$
and $N^i \equiv 2\ellc^{-1}$. 
We also define the (historical) dimensionless ratio
\be\label{Defsqrth}
\sqrt{h}\equiv\frac{\ells}{\ellc}  \quad \Rightarrow \quad A = \sqrt{h}\,\ells^{-1} \,.
\ee
The full set of Bianchi models considered in this work, as well as their underlying isotropic 3-spaces ${\cal M}^{(3)}$, are summarized in Table~\ref{TableType}. For future reference, note that models $\I$ and $\V$ can 
be obtained from models $\VIIo$ and $\VIIh$ in the limit $\ells\rightarrow\infty$.
\begin{table}[!htb]
	\begin{tabular}{c|cccc|c|c}
		Type & $A$ & $N^1$ & $N^2$ &$N^3$ & $\frac{a^2\Rspatial}{6}=K$ & ${\cal M}^{(3)}$ \\
		\hline
		${\rm I}$ & 0 & 0 & 0 & 0 & 0 & $\mathbb{E}^3$ \\
		${\rm V}$ & $\ellc^{-1} $& 0 & 0 & 0 & $-\ellc^{-2}$ & $\mathbb{H}^3$\\
		${\rm VII}_0$ & 0 & $\ells^{-1}$ & $\ells^{-1}$ & 0 & 0 & $\mathbb{E}^3$\\
		${\rm VII}_h$ & $\ellc^{-1}$ & $\ells^{-1}$ & $\ells^{-1}$ & 0 & $-\ellc^{-2}$ & $\mathbb{H}^3$\\
		${\rm IX}$ & 0 & $2\ellc^{-1} $ & $2\ellc^{-1}$& $2\ellc^{-1} $ & $+\ellc^{-2}$ & $\mathbb{S}^3$\\
	\end{tabular}
	\caption{Bianchi types considered in this work (first column), and their underlying maximally symmetric 3-spaces (last column), namely, the Euclidean ($\mathbb{E}^3$), hyperbolic ($\mathbb{H}^3$) and spherical ($\mathbb{S}^3$) spaces. For comparison, we also give (a combination of) the spatial Ricci scalar appearing 
	in Friedmann equations for each of the 3-spaces ${\cal M}^{(3)}$.\label{TableType}}
\end{table}

Next, we define a spacetime basis of invariant vector fields by choosing a set of
spatial vectors at a reference point, and Lie dragging them with the KVFs on a given spatial section. That is we define the vector fields on that section by the conditions
\begin{subequations}
\begin{align}
\gr{e} \cdot \gr{e}_i & =   0\,,\label{ndoteI} \\
[\gr{\xi}_i,\gr{e}_j] & = {\cal L}_{\gr{\xi}_i} \gr{e}_j = 0\,.\label{Liexie}
\end{align}
\end{subequations}
These spatial vectors are extended throughout the other spatial sections by demanding that
\be\label{Liene}
[\gr{e},\gr{e}_i] = {\cal L}_{\gr{e}} \gr{e}_i  = 0\,.
\ee
Finally, we join the unit normal vector $\gr{e}$ to the set $\{\gr{e}_i\}$ to obtain a spacetime basis
$\{\gr{e}_a\}$, with the understanding that $\gr{e}_0=\gr{e}$. From properties \eqref{DefKVF}-\eqref{Propertyspatial}, 
it is shown that 
\be
{\cal L}_{\gr{e}} \gr{\xi}_i = [\gr{e},\gr{\xi}_i] = 0\,,
\ee
that is, the normal vector $\gr{e}$ is also invariant under the action of the KVFs. The Jacobi identity applied to 
$\gr{e}$, $\gr{e}_i$ and $\gr{\xi}_j$ then shows that (\ref{Liene}) is consistent. Moreover, the conditions (\ref{Defnfromt}) and (\ref{Propertyunit}) imply that $\gr{e}$ is geodesic ($e^\mu \nabla_\mu e^\nu = 0$).

The commutator of the basis vectors $\gr{e}_i$ is another vector, and can thus be represented as a linear 
combination of the basis elements
\be\label{Defncommut}
[\gr{e}_i,\gr{e}_j] = \tilde{C}^k_{\,\,ij}\gr{e}_k\,,
\ee
where $\tilde{C}^k_{\,\,ij}$ are constants. Since we are still free to fix the orientation of the spatial basis $\{\gr{e}_i\}$ at 
any point $p$, we choose $\{\gr{e}_i\}_p =
\{\gr{\xi}_i\}_p$~\footnote{This choice agrees with the one made in PC10.}. However, 
since our Eq.~\eqref{DefC} differs with theirs by a minus sign, so does our Eq.~\eqref{DefCfore}. This then gives
\be\label{DefCfore}
[\gr{e}_i,\gr{e}_j] = - {C^k}_{ij} \gr{e}_k\,,
\ee
which can be checked by writing $\gr{e}_i=M^j_i\gr{\xi}_j$, where $M^j_i$ is a point-dependent matrix obeying $M^j_i(p)=\delta^i_j$, and using Eq.~\eqref{Liexie}.
We thus find
\begin{align}\label{ExplicitCommuteBasis}
\begin{split}
[\gr{e}_1,\gr{e}_2] & = - N^3\gr{e}_3\,, \\
[\gr{e}_1,\gr{e}_3] & = + N^2\gr{e}_2 -A\gr{e}_1\,, \\
[\gr{e}_2,\gr{e}_3] & = - N^1\gr{e}_1 -A\gr{e}_2\,.
\end{split}
\end{align}
Using again the Jacobi identity for $\gr{e}$, $\gr{e}_i$ and $\gr{e}_j$ one can show that
\be
{\cal L}_{\gr{e}}{C^i}_{jk} = 0\,,
\ee
that is, the constants of structure are really spacetime constants for this invariant basis. Note that our choice
of a time-invariant basis contrasts with the more popular choice of a tetrad basis, in which the constants of structure become 
time-dependent~\cite{Ellis1969,Ellis1998,Sung:2010ek}.

From the time-invariance property (\ref{Liene}) and from (\ref{DefCfore}), we infer that
\be\label{DefCABC}  
[\gr{e}_a,\gr{e}_b] = - {C^c}_{ab} \gr{e}_c\,,
\ee
where
\be
{C^0}_{i0} = {C^i}_{j0} = {C^0}_{ab} = 0\,,
\ee	
that is, the constants of structure vanish whenever one of the indices is $0$.

Next, we define the dual basis $\{\gr{e}^a\}$ to the basis $\{\gr{e}_a\}$ from the condition
\be\label{DefCoBasis}
e^a_\mu e_b^\mu = \delta^a_b\,.
\ee
from where it follows that $e^0_\mu=-\omega_{\mu}=-e_\mu$. From (\ref{Propertyunit}) and (\ref{ndoteI}) we deduce that in 
this dual basis the components of the metric satisfy $g_{00} =-1$ and $g_{0i}=0$. Since the metric has three spacelike KVFs, 
it can at most depend on $t$, so it is of the form
\be\label{Bianchi-general-metric}
\gr{g} = - \gr{e}^0 \otimes \gr{e}^0 + g_{ij}(t) \gr{e}^i \otimes \gr{e}^j\,.
\ee 
It is also convenient to define a spatial metric through
\be\label{Defh}
\gr{h} \equiv \gr{g} + \gr{e}^0\otimes \gr{e}^0 =  g_{ij}(t) \gr{e}^i \otimes \gr{e}^j\,,
\ee
such that $h_{ij}=g_{ij}$. From the covariant derivative $\nabla$ associated with the metric $\gr{g}$ we can define an induced covariant 
derivative $D$ associated with the induced spatial metric $\gr{h}$. For any spatial tensor $\gr{T}$ it is defined 
by~\cite{Tsagas:2007yx,Ellis1998}
\be\label{DefD}
D_\mu T_{\nu_1\dots \nu_n} \equiv h^\sigma_\mu h^{\lambda_1}_{\nu_1}\cdots h^{\lambda_n}_{\nu_n}
\nabla_\sigma T_{\lambda_1\dots\lambda_n}\,.
\ee

From the definition \eqref{DefCoBasis} and the property \eqref{DefCABC}, we
deduce that the constants of structure also satisfy the
property~\footnote{\cpsilent{Antisymmetrization on $n$ indices is defined with a
prefactor $1/n!$, that is $T_{[ij]} = (T_{ij}-T_{ji})/2$. However
there is no such prefactor in commutators.}}
\be\label{Cvsform}
{C^c}_{ab} = 2 e_a^\mu e_b^\nu \nabla_{[\mu} e^c_{\nu]}\,.
\ee
Next we introduce the connection coefficients through
\be
{\Gamma^c}_{ab} \equiv - e_a^\mu e_b^\nu \nabla_\mu e^c_\nu = e^c_\nu e_a^\mu \nabla_\mu e_b^\nu   \,.
\ee
Comparison with (\ref{Cvsform}) then shows that ${{C^c}_{ab} = -
  2{\Gamma^c}_{[ab]}}$. In particular 
\be
{\Gamma^0}_{ij} = - e_i^\mu e_j^\nu \nabla_\mu e^0_\nu = \tfrac{1}{2}{\cal L}_{\gr{e}_0} g_{ij} = \tfrac{1}{2}\dot g_{ij}\,,
\ee
which is related to extrinsic curvature $K_{\mu\nu} \equiv
h^\alpha_\mu h^\beta_\nu \nabla_\alpha e_{\beta}$ by
\be\label{DecKij}
K_{ij} = {\Gamma^0}_{ij} \equiv \tfrac{1}{3} \theta g_{ij} + \sigma_{ij}\,.
\ee
Here, we have separated its trace (proportional to the volume expansion $\theta$) from its traceless 
part (given by the expansion shear $\sigma_{ij}$). 

Using also that the connection is torsionless we can relate its components to the
constants of structure by~\footnote{$e_a^\mu \partial_\mu g_{bc} = \Gamma_{cab}+\Gamma_{bac}$ from metric compatibility, 
and $\Gamma_{c[ab]}=-C_{cab}/2$ from torsionless conditions. } 
\bea
\Gamma_{abc} &=& \frac{1}{2}\left[-e_a^\mu \partial_\mu
  g_{bc}+e_b^\mu \partial_\mu g_{ca}+e_c^\mu \partial_\mu g_{ab}\right.\nonumber\\
&&\qquad \left.+C_{acb} - C_{bac} + C_{cba}\right]\,,
\eea
where we have introduced the definitions
\be
\Gamma_{abc}\equiv g_{ad}{\Gamma^d}_{bc}\,, \quad C_{abc} \equiv g_{ad} {C^d}_{bc}\,.
\ee
For the spatial components we get simply
\be\label{GCIJK}
\Gamma_{ijk} = \tfrac{1}{2}\left[C_{ikj} - C_{jik} + C_{kji}\right]\,.
\ee
Since the Riemann tensor is associated with the connection and its derivatives, then from
(\ref{GCIJK}), this implies that the Riemann tensor associated with $\gr{h}$ can be
fully expressed in terms of the constants of structure ${C^i}_{jk}$
(see Appendix~\ref{1p3review}). The Gauss-Codazzi identity
\cite{Gourgoulhon:2007ue,xPand} allows to relate the Riemann tensor
associated with the metric $\gr{h}$ and its connection $D$ to the
spacetime Riemann tensor associated with $\gr{g}$ and connection
$\nabla$. The most general form of this identity is given by \eqref{EqGaussCodazzi}.

Finally, we would like to stress the importance of the invariant basis~\eqref{Liexie}. Indeed,
the definition of a homogeneous tensor as any tensor $\gr{T}$ such that ${\cal L}_{\gr{\xi}_i}\gr{T} = 0$
is natural, since in this basis $\gr{T}$ has constant components. In particular, the quantities ${C^i}_{jk}$ 
can be interpreted as the components of an underlying homogeneous tensor fields $\gr{C}={C^i}_{jk} \gr{e}_i \otimes \gr{e}^j \otimes
\gr{e}^k$.

\subsection{Conformal parameterization}

Using the time coordinate $t$ introduced in~\eqref{Defnfromt}, the metric of a general Bianchi spacetime reads
\be\label{DefB-metric}
\gr{g}^{\rm Bianchi} = - \dd t \otimes \dd t + a^2(t) \gamma_{ij}(t) \gr{e}^i \otimes \gr{e}^j\,,
\ee
At this point, it is also convenient to define a conformal time by
\be\label{conformal-time}
a(\eta) \dd \eta \equiv \dd t\,,
\ee
which implies that, in conformal time, ${e^\mu=a^{-1}\delta^\mu_0}$. The volume expansion $\theta$ is related to the conformal Hubble 
rate by
\be
\HH \equiv \frac{a'}{a} = \frac{a\theta}{3}\,,
\ee
where, throughout this work, a prime indicates a derivative with
respect to $\eta$. The conformal spatial metric $\gamma_{ij}$ is
defined by
\be\label{Defgammaij}
h_{ij} = a^2 \gamma_{ij}
\ee
such that the Bianchi metric takes the form \eqref{DefB-metric}. Note
that the derivative \eqref{DefD} can also be considered as being
associated with $\gamma_{ij}$.

The conformal shear $\hat{\sigma}_{ij}$ is defined by
\be\label{sigma-hat}
\hat\sigma_{ij} \equiv \frac{1}{2}\gamma_{ij}' \,.
\ee
The components of the constants of structure are related to their conformal counterparts by
\bea\label{DefHatNHatA}
N^{ij} &=& a^{-3}\hat N^{ij}\,,\; N_{ij} = a^1 \hat N_{ij}\,,\; N_i^j = a^{-1} \hat N_i^j\,,\nonumber\\
A_i &=& \hat A_i\,, \quad A^i = a^{-2} \hat A^i\,.
\eea
We also have that
\be
\sigma_{ij} =a \hat \sigma_{ij} \,,\qquad \sigma_i^j = a^{-1} \hat \sigma_i^j\,.
\ee
This means that in practice the indices of $\hat \sigma_{ij}$, $\hat N^{ij}$ and $\hat A_i$ are raised and lowered by $\gamma_{ij}$ and $\gamma^{ij}$, 
whereas those of $\sigma_{ij}$, $N^{ij}$ and $A_i$ are raised and
lowered by $a^2 \gamma _{ij}$ and $a^{-2 }\gamma^{ij}$. The Levi-Civita tensor is also
decomposed as $\epsilon_{ijk} = a^{3} \hat \epsilon_{ijk}$ with
$\hat{\epsilon}_{123} = 1$ such that the combination $\hat
\epsilon_{ijl} \hat N^{lk} $ in the decomposition \eqref{DefNA} is equal to $\epsilon_{ijl} N^{lk}
$.

Since the constants of structure are constant, these definitions ensure that the conformal $\hat N^{ij}$ and $\hat A_i$ and their related forms with different index placements 
are constant. 

\subsection{Stress-energy tensor}\label{SecTmunu}

The stress-energy tensor of a fluid with energy density $\rho$, pressure $p$ 
and anisotropic stress $\pi_{\mu\nu}$ is
\be\label{tmunu-pf}
T_{\mu\nu} = (\rho+p) u_\mu u_\nu + p g_{\mu\nu} + \pi_{\mu\nu}\,,
\ee
where $u^\mu$ is the (timelike) fluid four-velocity. For simplicity we
assume no anisotropic stress, although its inclusion is
straightforward. Homogeneity of Bianchi space-times implies that energy density and pressure
depend only on time, and therefore we use the notation $\bar \rho$ and $\bar p$ for the fluid content of Bianchi universes to stress this fact. 
We stress that the vector $u^\mu$ is not necessarily parallel to $e^\mu$, since a homogeneous boost of the fluid is allowed for tilted Bianchi models. Actually, 
the fluid's four-velocity can be decomposed into components parallel and orthogonal to $e^\mu$ so that in the invariant basis we have
\be
u^\mu = \Gamma(e^\mu + v^\mu),\quad\textrm{with}\quad\Gamma = \frac{1}{\sqrt{1-v_i v^i}}\,,
\ee
where $v^\mu e_\mu =0$. If this homogeneous velocity is not curl-free, then this corresponds
to a global rotation of the fluid~\cite{Collins1973a,Collins1973b}.

Moving forward, we introduce comoving components for the velocity  as follows
\be
v_i = a\hat{v}_i\,,\quad v^i = a^{-1}\hat{v}^i\,,
\ee
such that the velocity has components
\be\label{Ucomponents}
u^\mu = \frac{\Gamma}{a}(1,\hat v^i),\quad u_\mu = a \Gamma (-1,\hat v_i)\,.
\ee
Finally, the stress-energy tensor components are
\begin{align}\label{Tij}
\begin{split}
-T^\eta_\eta & = T_{\mu\nu} e^\mu e^\nu=\bar \rho+(\bar \rho+\bar p)(\Gamma^2-1)\,,\\
T^\eta_{i} & = (\bar \rho+\bar p)\Gamma^2 \hat v_i\,,\\
T^i_{j} & = [(\bar \rho+\bar p)\Gamma^2 \hat v^i \hat v_j + \bar p \delta^i_{j}]\,.
\end{split}
\end{align}

Because we allow for a tilt, $T_{\mu\nu}$ will not look like~\eqref{tmunu-pf} with $u_\mu \to e_\mu$ for observers following the congruence 
defined by $e^\mu$. In fact, from \eqref{Tij} we check that for those observers the fluid will present an effective momentum density and anisotropic 
stress. It then follows from Einstein equations that these components will source spatial anisotropies even in the case of a perfect fluid.

\subsection{Einstein equations}

The time-time component of Einstein equation follows by projecting $G_{\mu\nu} = 8\pi G T_{\mu\nu}$ with $e^\mu$. 
Defining $\kappa = 8\pi G$ and using the relation \eqref{GaussCodazziR} we get 
\be\label{FirstFriedEq}
{\cal H}^2 
 + \frac{a^2}{6}\Rspatial - \frac{1}{6}\hat\sigma^2 = \frac{\kappa a^2}{3}T_{\mu\nu} e^\mu e^\nu
\ee
where $\hat{\sigma}^2 \equiv \hat{\sigma}_{ij}\hat{\sigma}^{ij}$ and the spatial Ricci scalar is given in terms of constants of
structure to be [see \eqref{EqR3} with definitions \eqref{DefHatNHatA}]
\be
a^2\,\Rspatial  = -6\hat{A}_i \hat{A}^i -\hat{N}_{ij}\hat{N}^{ij}+\frac{1}{2}(\hat{N}_i^i)^2\,.
\ee
The evolution of the expansion rate is given by the Raychaudhuri equation \eqref{Raychaudhuri}
\be
3\HH'+ \hat{\sigma}^2 = - a^2\kappa\left(T_{\mu\nu}e^\mu e^\nu+\frac{1}{2}T\right)\,,
\ee
and the dynamics of the shear comes from the traceless part of the spatial Einstein equation
\bea
(\hat \sigma_{ij})' + 2\HH \hat \sigma_{ij}  &=& \hat N^k_k \hat
N_{\langle i j\rangle} 
-2 \hat N_{k \langle i} \hat N^k_{j\rangle} \\
&& + 2 \hat A^k\hat \epsilon_{kl\langle j} \hat N_{i
  \rangle}^{\,l}+ \kappa T_{\langle ij \rangle}\,, \nonumber
\eea
where $ T_{\langle ij \rangle} = a^2 (\bar \rho+\bar p)\Gamma^2 \hat v_{\langle i}\hat v_{j \rangle}$.

Lastly, there is a constraint equation following from $G^\eta_{i}=\kappa
T^\eta_{i}$, which is 
\be
P_i  \equiv \kappa a^2 T^\eta_{i} = 3 \hat{A}^j \hat \sigma_{ji} + \hat{\epsilon}_{ijk} \hat{\sigma}^{jl}\hat N_l^k\,.
\ee
This is known as the tilt constraint and by definition
\be\label{DefTilt}
P_i = \kappa a^2 \Gamma^2 (\bar \rho+\bar p) \hat{v}_i\,.
\ee
But this is what is used to deduce $\hat{v}_i$ that we must then replace in the shear equation because it is inside $T_{ij}$.

\subsection{Fluid equations}

As usual, fluid equations follow from the covariant divergence of the stress-energy tensor. The conservation equation for energy density is just
\be
\bar \rho' + (\bar \rho+\bar p)[3 \HH + (\ln \Gamma)' + D_i \hat v^i] = 0\,,
\ee
and from \eqref{UsefulDVDT}, we get the velocity divergence
\be
D_i \hat v^i = -2 \hat A_i \hat v^i\,.
\ee
When linearizing in the velocity $\hat v_i$, the term $(\ln \Gamma)'$ will behave as a second order quantity, and thus will
not contribute.

The Euler equation is formally just
\be
[a^4 (\bar \rho+\bar p)\Gamma^2 \hat v_i]' + a^4 (\bar \rho+\bar p)\Gamma^2 D_j (\hat v^j \hat v_i) = 0\,.
\ee
Separating the trace and traceless parts, the second term is handled
using \eqref{UsefulDVDT}. However, when linearizing in the velocity
$\hat v_i$ it vanishes hence the Euler equation reduces to
\be
[a^4 (\bar \rho+\bar p) \hat v_i]' \simeq 0\,.
\ee

\section{SVT modes in FLRW}\label{Sec4-SVT-FLRW}

We now give a brief review of the background geometry of FLRW models and the mathematics behind the standard Scalar-Vector-Tensor (henceforth SVT) decomposition of perturbative modes. This will be needed when comparing perturbations of the FLRW metric with spatial anisotropies in the homogeneous limit. 

\subsection{Background FLRW cosmology}

Given a cosmic time $t$ which allows us to split spacetime into space and time, all FLRW metrics can be written in the form
\be\label{DefFL}
\gr{g}^{\rm FLRW} = -  \dd t \otimes \dd t + a^2(t) \gr{g}^{\rm MS}
\ee
where $a(t)$ is the scale factor of the expansion and $\gr{g}^{\rm MS}$ is the metric of maximally symmetric spaces, described by
\be
\gr{g}^{\rm MS} =\left[\dd  \chi^2 + r^2(\chi) \dd^2 \Omega\right]\,.\label{dsspat}
\ee
Here, $\dd^2 \Omega \equiv \dd \theta^2 + \sin^2 \theta \dd \phi^2$ is the standard line element on the 2-sphere and the function $r(\chi)$ is given by
\be\label{Defrchi}
r(\chi)=\begin{cases}
\ellc \sinh(\chi/\ellc)\,, & K<0\,,\\
\chi \,, & K=0\,,\\
\ellc \sin(\chi/\ellc)\,, & K>0\,.
\end{cases}
\ee

The curvature parameter $K$ differentiates between open ($K<0$), flat ($K=0$) and closed ($K>0$) spatial sections. 
It is related to the curvature radius $\ell_c$ of the spatial sections
by $\ell_c=1/\sqrt{|K|}$. It is also useful to introduce the quantity
\be\label{dimensionlessK}
{\cal K} \equiv K/|K| = K \ellc^2\,,
\ee
which is either $-1$, $0$ or $+1$ for open, flat and closed cases respectively.

\subsection{Linear perturbation theory}\label{sec:lpt}

To complete the matching between Bianchi anisotropies and FLRW
perturbations, we will also need equations from perturbation theory in
synchronous gauge (usually fixed completely with an additional comoving condition
on cold dark matter) and in conformal time, which will now be briefly
summarized. More details can be found in Refs.~\cite{Ma1995,pubook,Malik:2008im}. 

By definition, only the spatial part of the metric is perturbed in synchronous gauge. In the SVT decomposition, such perturbations can be parameterized as follows: 
\be\label{SVT-metric}
\delta g_{ij} = 2a^2\left[-\phi g^{\rm MS}_{ij}+\hatD_{ij}\psi + \hatD_{(i}E_{j)} + E_{ij}\right],
\ee
where $\hatD_i$ is the covariant derivative compatible with $g^{\rm MS}_{ij}$ and 
\be\label{DefDelta}
\hatD_{ij} \equiv \left(\hatD_i \hatD_j - \frac{g^{\rm
      MS}_{ij}}{3}\Delta \right)\,,\quad\Delta \equiv \hatD_i \hatD^i\,.
\ee
At linear order, the perturbed components of the Einstein tensor can also be decomposed into scalar, vector and tensor modes~\cite{pubook}. Working in conformal time (recall Eq.~\eqref{conformal-time}), such components are given by
\begin{widetext}
\paragraph*{Scalar modes:} 
\beas
a^2\delta G^{\eta}_{\eta} & = & 6\HH \phi'- (\Delta+3K)2\phi -
\hatD^i \hatD^j \hatD_{ij}\psi\,, \\
a^2\delta G^{\eta}_{i} & = & -2\hatD_i\phi'-\hatD^j \hatD_{ij}\psi'\,, \\
a^2\delta G^{i}_{j} & = &  \left[\partial_\eta^2 + 2 \HH \partial_\eta
  +\frac{1}{3} (\Delta -6K) \right] \hatD^i_{\,j} \psi+\hatD^i_{\,j}\phi\\\nonumber
&+& 2\delta^i_j\left[\phi''+2\HH\phi'-\frac{1}{3}(\Delta+3K)\phi-\frac{1}{6}\hatD^k \hatD^l \hatD_{kl}\psi\right]\,.
\eeas

\paragraph*{Vector modes:}
\beas
a^2\delta G^{\eta}_{i} & = & -\frac{1}{2}\left(\Delta+2K\right)E'_i\,, \\
a^2\delta G^{i}_{j} & = & g_{\rm MS}^{ik}\hatD_{(k}\left[E''_{j)}+2\HH E'_{j)}\right]\,.
\eeas
\paragraph*{Tensor modes:}
\be
a^2\delta G^{i}_{j} = E''^i_{j}+2\HH E'^i_{j} - (\Delta-2 K)E^i_{j}\,.
\ee
\end{widetext}

We also give the Ricci scalar associated with the spatial metric. For isotropic backgrounds and at linear order, it is sourced only by scalar perturbations:
\be\label{dRtroisFL}
a^2 \delta(\Rspatial) = 4(\Delta+3K) \phi + \frac{4}{3}\Delta(\Delta + 3 K) \psi\,.
\ee

In order to proceed with the identification, we will also need to perturb the energy-momentum tensor, here taken to be that of a perfect fluid for simplicity. As it turns out, the linearized tensor in synchronous gauge is exactly what one would obtain by setting $\Gamma=1$ in~\eqref{Tij} and linearizing $\rho$ and $p$ around their background values (denoted below by an overbar). This leads to
\bea\label{Tij2}
-T^\eta_\eta& =  &\bar \rho + \delta \rho\,,\\
T^\eta_{i} &=& (\bar \rho+\bar p)\hat v_i\,,\\
T^i_{j}&=& (\bar p + \delta p) \delta^i_{j}\,.
\eea
Note that the velocity $\hat{v}_i$ is considered as a first-order perturbation, and as such it is split into scalar and vector parts as
\be
\hat{v}_i=\hatD_i\hat{v}+\hat{V}_i\,,\quad \hatD^i \hat{V}_i = 0\,.
\ee

We thus have everything needed to write the perturbed Einstein equations. The first of them follows from the time-time component, and corresponds to the (perturbed) Friedmann equation:
\be
-6\HH \phi' +  (\Delta+3K)2\phi + \hatD^i \hatD^j \hatD_{ij}\psi = \kappa a^2 \delta \rho\nonumber
\ee
Then, we have the trace-free components of the space-space Einstein equations. These are independently given for scalar
\be\label{trace-free-efe}
\left[\partial_\eta^2 + 2 \HH \partial_\eta +\frac{1}{3} (\Delta
-6K) \right] \hatD^i_{\,j} \psi+\hatD^i_{\,j}\phi = 0\,,
\ee
vector
\be
\hatD_{(i} E''_{j)}+2\HH \hatD_{(i} E'_{j)} = 0,\label{VectorEqFL}
\ee
and tensor modes
\be
E''^i_{j}+2\HH E'^i_{j} - (\Delta-2 K)E^i_{j} = 0.\label{TensorEqFL}
\ee
Next, we have two constraint equations for the scalar and vector modes of the velocity perturbation. These are given by
\begin{subequations}
\begin{align}
-2\hatD_i\phi'-\hatD^j \hatD_{ij}\psi' & = \kappa a^2 (\bar \rho+\bar p)\hatD_i \hat v\,,\slabel{1st-v-constraint} \\
-\frac{1}{2}\left(\Delta+2K\right)E'_i & = \kappa a^2 (\bar \rho+\bar p)\hat V_i\,.\slabel{2nd-v-constraint} 
\end{align}
\end{subequations}

Finally, from the conservation of the energy-momentum tensor we have the Euler equation
\be\label{EulerEq-FL}
[a^4 (\bar \rho+\bar p)\hat v_i]'=-a^4 \partial_i \delta p\,,
\ee
and the conservation equation
\be\label{ConservEq-FL}
\delta\rho' + (\delta\rho+\delta p)3\HH = (\bar\rho+\bar p)(3 \phi' - \hatD_i \hat v^i).
\ee

\subsection{Harmonics}\label{subsec:laplace-eigen}

Since perturbative modes evolve independently for linear perturbations, it is convenient to expand the metric perturbation on 
a basis of tensor harmonics. In this section we summarize how these harmonics are built (see \cite{Pitrou:2019ifq} for more details).

Let us first introduce the usual orthonormal spatial basis associated with spherical coordinates
\begin{align}
\begin{split}
\gr{n} & = \partial_\chi\,,\\
\gr{n}_\theta & = r^{-1}(\chi)\partial_\theta\,,\\
\gr{n}_\phi & = r^{-1}(\chi)\csc\theta \partial_\phi\,.
\end{split}
\end{align}
They allow us to introduce the standard helicity (vector) basis,
\be\label{npm}
\gr{n}_\pm \equiv \frac{1}{\sqrt{2}}\left(\gr{n}_\theta \mp \ii  \gr{n}_\phi\right) \,,
\ee
which is in turn used to defined an extended helicity (tensor) basis~\cite{Pitrou:2019ifq}
\be\label{Defns}
\hat{n}_{\pm s}^{i_1\dots i_j}(\gr{n}) \equiv n_\pm^{\langle i_1}\dots n_\pm^{i_s}
n^{i_{s+1}}\dots n^{i_j\rangle}\,.
\ee
In what follows, we shall occasionally use a multi-index notation
\be
I_j \equiv i_1 \dots i_j\,,
\ee
such that the helicity basis is written simply as $\hat{n}^{\pm s}_{I_j}$ or as $\hat{n}_{\pm s}^{I_j}$. 

The generalized helicity basis has a series of important properties which are collected in Ref.~\cite{Pitrou:2019ifq}. For our present purposes, we stress that it is both a complete basis for symmetric and trace-free tensors, as well as a natural basis for separating the angular from the radial dependence of spin-valued tensors.

Cosmological perturbations can be expanded in a basis of spatial eigenfunctions of the Laplace operator $\Delta$. Any tensor-valued perturbation $Q^{(jm)}_{I_j}(\gr{x};k)$ satisfying~\cite{Pitrou:2019ifq}
\be\label{laplace-beltrami-k}
[\Delta +k^2-K(j-|m|)(j+|m|+1)]Q^{(jm)}_{I_j} = 0\,,
\ee
will be loosely called an harmonic. Here, $j$ represents the tensorial
rank of the harmonic, and $m$ the rank of the primitive tensor from
which this harmonic is derived. Thus, for instance, $Q^{(2,0)}_{ij}$
is a rank-2 tensor derived from (two derivatives of) a scalar
function. The case $j=|m|$ represents a ``fundamental'' harmonic, in
the sense that it is not derived from lower-rank tensors, and it is
also divergence-less. Harmonics with $j>|m|$ are derived from the
fundamental ones. In what follows, it will be convenient to give the expression above in terms of the mode $\nu_m$ 
defined as
\be\label{DefNu}
\nu^2_m \equiv k^2 + (1+|m|)K
\ee
in terms of which~\eqref{laplace-beltrami-k} becomes
\be\label{laplace-beltrami}
\{\Delta +\nu^2-K[(1-|m|)(1+|m|)+j(j+1)]\}Q^{(jm)}_{I_j} = 0\,.
\ee

Let us summarize the main definitions and results of \cite{Pitrou:2019ifq}
in which these harmonics and their decomposition into angular and
radial functions are discussed. A point $\gr{x}$ in space is specified by its distance ($\chi$) and direction ($\gr{n}$) from a given origin. However, cosmological perturbations are characterized independently by their spatial ($\gr{x}$) and angular ($\gr{n}$) dependencies, which requires the use of both orbital and spin eigenfunctions~\cite{TAM1,TAM2}. Since what is observed is the total angular dependence, one introduces a set of total angular momentum normal modes~\cite{TAM2} which splits perturbations in their effective radial and angular dependencies:
\be\label{DefGsjm}
{}_s G_\ell^{(jm)}(\chi,{\gr{n}};\nu) \equiv c_\ell \,\,{}_s
 \alpha_\ell^{(jm)}(\chi,\nu) \,{}_s Y_{\ell}^{m}({\gr{n}})\,,
\ee
where
\be\label{Defcl}
c_\ell \equiv \ii^\ell \sqrt{4\pi(2\ell+1)}\,,
\ee
${}_sY_\ell^m(\gr{n})$ are spherical harmonics of spin $s$ and
${}_s\alpha_\ell^{(jm)}$ are radial functions. These radial functions
are zero whenever one of the following conditions is violated:
\be
j \geq {\rm max}(|m|,|s|)\,,\quad \ell\geq {\rm max}(|m|,|s|)\,.
\ee
Moreover, these functions are conventionally normalized at origin through the condition 
\be\label{AlphaChiZero}
\left.{}_s \alpha_\ell^{(jm)}\right|_{\chi=0} = \frac{1}{2j+1} \delta_{\ell j}\,.
\ee
It is also convenient to decompose the radial functions into even (electric) and odd (magnetic) types as
\be\label{defeb}
{}_{\pm s} \alpha_{\ell}^{(jm)} = {}_{s}\epsilon_{\ell}^{(jm)}
\pm \ii \;{}_s\beta_{\ell}^{(jm)} \,.
\ee
Formal expressions and identities satisfied by the radial functions are collected in~\cite{Pitrou:2019ifq}.

Next we introduce the harmonics $Q^{(jm)}_{I_j}$ in the $\ell$-representation. These are constructed by a simple combination of the normal modes with the generalized helicity basis as
\be\label{Qljm}
{}^\ell Q^{(jm)}_{I_j} \equiv \sum_{s=-j}^j {}_s g^{(jm)}{}_sG_\ell^{(jm)}(\chi,\gr{n};\nu) \hat n^s_{I_j}(\gr{n})\,,
\ee
where ${}_s g^{(jm)}$ are numerical coefficients introduced in \cite{Pitrou:2019ifq}. Overall, when following reference~\cite{Pitrou:2019ifq}, the reader should make use of the replacements $\chi \to \chi/\ellc$, $k \to k \ellc $ and $\nu \to \nu \ellc$.

\subsection{Plane-waves and pseudo plane-waves}\label{subsec:planewave}
Finally, we build generalized plane-wave harmonics from summation over $\ell$:

\be\label{DefQfromsuml}
Q^{(jm)}_{I_j}\equiv \sum_{\ell\geq |m|} \frac{\myzeta_\ell^m}{\myzeta_{j}^{m}}\times{}^\ell Q^{(jm)}_{I_j}(\chi,\gr{n};\nu)
\ee
where $\myzeta_\ell^m$ are coefficients that can be fixed up to an overall arbitrary constant (see Appendix~\ref{Appzeta}). Likewise, a plane-wave normal mode is
\be\label{GjmfromGljm}
{}_s G^{(jm)}\equiv \sum_{\ell\geq |m|} \frac{\myzeta_\ell^m}{\myzeta_{j}^{m}}\times{}_s G^{(jm)}_{\ell}(\chi,\gr{n};\nu) \,,
\ee
such that \eqref{Qljm} also holds without the $\ell$ indices, that is after summation on $\ell$.
Usual plane-waves correspond to the case ${\myzeta_\ell^m = {\rm
    const.}}$, but the previous definitions allow for pseudo plane
waves when ${\myzeta_\ell^m \neq {\rm const}}$. \cpsilent{The plane-waves
harmonics were built using the zenith direction as a reference, but we
can rotate them so as to define harmonics with respect to the wave
vector $\gr{\nu} = \nu \hat{\gr{\nu}}$, as detailed in section 6.2 of
\cite{Pitrou:2019ifq}.} The perturbations defined in \eqref{SVT-metric}
are expanded on the basis~\eqref{DefQfromsuml} as
\begin{align}
\begin{split}\label{ReplacementRuleBardeen}
\phi &\rightarrow \int \frac{\dd^3 \gr{\nu}}{(2 \pi)^3} H_S^{(0)}(\gr{\nu},\eta)\,  Q^{(0,0)}(\gr{\nu})\,,  \\
\hatD_{ij} \psi &\rightarrow \int \frac{\dd^3 \gr{\nu}}{(2 \pi)^3} H_T^{(0)}(\gr{\nu},\eta)\,  Q^{(2,0)}_{ij}(\gr{\nu})\,, \\
\hatD_{(i} E_{j)} &\rightarrow \sum_{m=\pm 1}\int \frac{\dd^3 \gr{\nu}}{(2 \pi)^3} H_T^{(m)}(\gr{\nu},\eta)\,  Q^{(2,m)}_{ij}(\gr{\nu})\,,\\
E_{ij} &\rightarrow \sum_{m=\pm 2}\int \frac{\dd^3 \gr{\nu}}{(2 \pi)^3} H_T^{(m)}(\gr{\nu},\eta)\,  Q^{(2,m)}_{ij}(\gr{\nu})\,.
\end{split}
\end{align}

For applications related to observations, most notably those of CMB, it is convenient to work with the propagating direction $\bar{\gr{n}}$ related to the line-of-sight direction ${\gr{n}}$ through
\be
\bar{\gr{n}} \equiv -\gr{n}\,.
\ee
We can also define harmonics, with related normal modes and radial functions, associated with this convention. They are trivially related
to the previous ones by
\be
\overline{Q}^{(jm)}_{I_j}(\chi,-\bar{\gr{n}};\nu)  = (-1)^j
\times Q^{(jm)}_{I_j}(\chi,{\gr{n}};\nu)\,.\nonumber
\ee
The associated normal modes ${}_s \overline{G}^{(jm)}$ and radial
functions ${}_{\pm s} \overline{\alpha}_{\ell}^{(jm)} $ are related to
the previous ones as detailed in \S~7.1 of \cite{Pitrou:2019ifq}.

\subsection{Super-curvature modes}\label{SuperCurvature}
Quite generally, square-integrable cosmological perturbations can be constructed by superposing tensor harmonics characterized by ${\nu\ellc\geq 0}$. For closed spaces one further requires $\nu\ellc-1$ to be an integer larger or equal to $|m|$, see e.g. \cite{Pitrou:2019ifq}. In the case of open spaces ($K=-\ellc^{-2}$), this requires $(k\ellc)^2 \geq (1+|m|)$ (see~\eqref{DefNu}). While the inclusion of modes in this range is enough to describe perturbations that decay at infinity, it has been argued~\cite{Lyth1995} that the most general Gaussian perturbations also require the inclusion of modes having
\be\label{Cond1}
-1\leq (\nu \ellc)^2\leq 0\,.
\ee
In the scalar case ($m=0$) this corresponds to $0 \leq k^2 \leq
\ellc^{-2}$. For this reason, these are known as super-curvature modes.

Super-curvature harmonics are not square-integrable. Since they
correspond to purely imaginary $\nu$, they can be defined from
analytic continuation of the radial functions ${}_s
\alpha^{(jm)}_\ell$ of the usual harmonics (i.e., those with $\nu\geq
0$). In~\cite{Lyth1995} only the scalar harmonics were considered, but
the procedure of analytic continuation can be followed for all types
of harmonics. In fact the analytic continuation is not restricted to
\eqref{Cond1} but can be extended at least to the whole subset of the complex plane defined by
\be
 -1 \leq {\rm Im}(\nu \ellc) \leq 1\,\,\Rightarrow \,\,{\rm Re}[(\nu \ellc)^2] \geq -1\,.
 \ee
For that, one only needs to know how to formally build the radial functions -- see~\cite{Pitrou:2019ifq}. Hereafter, we call the case $(\nu \ellc)^2 = -1$ the \emph{maximal super-curvature mode} as it corresponds to $k=0$ for scalar harmonics.
  
Given that $\nu$ can be complex, the electric and magnetic parts of the
radial functions are not necessarily real-valued functions. In
particular the relation ${}_s\alpha_\ell^{(jm)\star}(\chi,\nu) =
{}_{-s}\alpha_\ell^{(jm)}(\chi,\nu)$, which holds in the flat case, is
not valid anymore and we must rely on Eq. (3.35) of \cite{Pitrou:2019ifq}.

\section{Linearization of Bianchi space-times }\label{sec:bianchi-modes}

Our goal now is to derive the linearized dynamical equations for the Bianchi models in Table~\ref{TableType}, which will be ultimately matched to the Einstein and fluid equations presented in the last section. Such matching requires contrasting Eqs.~\eqref{DefFL} and~\eqref{DefB-metric}, which in turn depends on the knowledge of the co-basis $\gr{e}^i$ in a given coordinate system. Since the co-basis is the dual to the invariant basis $\gr{e}_i$, which is in turn defined by the KVFs through Eq.~\eqref{Liexie}, we start this section by recalling the KVFs and invariant basis for the selected Bianchi models.

\subsection{KVF and invariant basis}\label{KVF-and-invbasis}

An ingenious method to find the KVFs of Bianchi models with FLRW limit was proposed in PC10, and can be summarized as follows: starting from a maximally symmetric space, one identifies its translational ($\gr{T}_i$) and rotational ($\gr{R}_i$) KVFs, as well as their commutators. Next, one 
looks for constants $\rho_i^j$ such that the newly defined vectors
\be
\gr{\xi}_i \equiv \gr{T}_i + \rho_i^j\gr{R}_j\,,
\ee
satisfy Eq.~\eqref{DefC}. From these vectors one then obtains the invariant basis $\gr{e}_i$ (through Eq.~\eqref{Liexie}) and their associated co-basis $\gr{e}^i$ which, once multiplied by a time-dependent tensor $\gamma_{ij}(t)$, and following the prescription of~\S\ref{subsec:general-construction}, leads to the metric~\eqref{DefB-metric}. 

The beauty of this method is that one naturally sees which Bianchi models can emerge from a given maximally symmetric space. Moreover, coordinate systems for the KVFs are naturally inherited from the coordinate systems of the underlying symmetric space. We now summarize these vectors and their associated invariant basis. Their co-basis can then be obtained from the prescription given in Appendix~\ref{AppCoBasis}.
 
\subsubsection{Models $\I$ and $\VIIo$}

Bianchi models $\I$ and $\VIIo$ are the only models emerging from flat Euclidean space. As such, their Killing vectors and invariant fields can be expressed in terms of natural (cartesian) coordinates of the flat Euclidean metric:
\be
\gr{g}^{\rm MS} = \dd x^2 + \dd y^2 + \dd z^2\,.
\ee 
The set of KVFs and invariant basis for model $\I$ corresponds to simple spatial translations:
\be\label{BI-cobasis}
\gr{\xi}^{(\I)}_i = \partial_i\,,\quad \gr{e}^{(\I)}_i = \partial_i\,.
\ee
It is trivial to check that $\gr{\xi}^{(\I)}$ satisfies~\eqref{DefC} with ${C^k}_{ij}=0$, and that~\eqref{Liexie} is satisfied automatically. 

The isometries of model $\VIIo$ correspond to two simple translations, and a translation followed by a rotation. Choosing this rotation to be around the $z$-axis, we then have for the KVFs:
\beas
\gr{\xi}^{(\VIIo)}_1 & =& \partial_x\,,\\
\gr{\xi}^{(\VIIo)}_2 & =& \partial_y\,,\\
\gr{\xi}^{(\VIIo)}_3 & =& \partial_z-\ells^{-1}(x\partial_y-y\partial_x)\,.
\eeas
One can check that these vectors satisfy~\eqref{ExplicitCommuteXi} with $N^1=N^2=\ells^{-1}$ and $A=0$. 
The invariant basis which solves~\eqref{Liexie} is:
\be\label{OneToSevenZero}
\gr{e}_i^{(\VIIo)} = M_i^{\,\,j}  \gr{e}_j^{(\I)}\,,
\ee
where $M_i^{\,\,j}$ are the components of the rotation matrix around the $z$-axis by the angle $z/\ells$:
\be\label{spiral-matrix}
\gr{M}=\left(\begin{array}{ccc}
	\cos (z/\ell_{s}) & -\sin (z/\ell_{s}) & 0\\
	\sin (z/\ell_{s}) & \cos (z/\ell_{s}) & 0 \\
	0 & 0 & 1
\end{array}\right)\,.
\ee
Since rotation matrices are orthogonal, we also have $M^i_{\,\,j} = M_i^{\,\,j}$ such that for the co-basis 
\be\label{FiveToSevenh2}
\gr{e}^i_{(\VIIo)} = M^i_{\,\,j} \gr{e}^j_{(\I)}\,.
\ee

\subsubsection{Models $\V$ and $\VIIh$}

These are the two models emerging from a maximally symmetric open space (i.e., a space with negative curvature). A possible coordinate system for the KVFs and invariant basis are spherical hyperbolic coordinates $(\chi,\theta,\phi)$, in terms of which the metric of the underlying space writes 
\be\label{openmetricspherical}
\gr{g}^{\rm MS} = \dd \chi^2 + \ellc^2 \sinh^2(\chi/\ellc) \dd^2 \Omega\,.
\ee
In these coordinates, the KVFs and invariant basis for model $\V$ are given respectively by
\begin{subequations}
\begin{align}\label{teste}
\gr{\xi}^{(\V)}_1 & = \sin\theta\cos\phi\partial_\chi \\
& +\ellc^{-1} [\cos\theta\coth(\chi/\ellc)-1]\cos\phi\partial_\theta \nonumber\\
& + \ellc^{-1} [\cot\theta-\coth(\chi/\ellc)\csc\theta]\sin\phi\partial_\phi\,,\nonumber\\
\gr{\xi}^{(\V)}_2 & = \sin\theta\sin\phi\partial_\chi \\
& +\ellc^{-1} [\cos\theta\coth(\chi/\ellc)-1]\sin\phi\partial_\theta\nonumber\\
& +\ellc^{-1}
  [\coth(\chi/\ellc)\csc\theta-\cot\theta]\cos\phi\partial_\phi\,, \nonumber\\
\gr{\xi}^{(\V)}_3 & = \cos\theta\partial_\chi - \ellc^{-1}\coth(\chi/\ell_c)\sin\theta\partial_\theta\,.
\end{align}
\end{subequations}
and
\begin{align*}
\gr{e}^{(\V)}_1 & = \frac{\sin\theta\cos\phi}{\cosh(\chi/\ellc)-\cos\theta\sinh(\chi/\ellc)}\partial_\chi\\
& + \frac{\ellc^{-1}[\cos\theta\coth(\chi/\ellc)-1]}{\cosh(\chi/\ellc)-\cos\theta\sinh(\chi/\ellc)}\cos\phi\partial_\theta\\
& -\ellc^{-1}\sin\phi\csc\theta\csch(\chi/\ellc)\partial_\phi\,, \\
\gr{e}^{(\V)}_2 & = \frac{\sin\theta\sin\phi}{\cosh(\chi/\ellc)-\cos\theta\sinh(\chi/\ellc)}\partial_\chi\nonumber  \\
& + \frac{\ellc^{-1}[\cos\theta\coth(\chi/\ellc)-1]}{\cosh(\chi/\ellc)-\cos\theta\sinh(\chi/\ellc)}\sin\phi\partial_\theta\\
& +\ellc^{-1}\cos\phi\csc\theta\csch(\chi/\ellc)\partial_\phi\,,\\
\gr{e}^{(\V)}_3 & = \frac{\cos\theta\cosh(\chi/\ellc)-\sinh(\chi/\ellc)}{\cosh(\chi/\ellc)-\cos\theta\sinh(\chi/\ellc)}\partial_\chi\\
& - \frac{\ellc^{-1}\csch(\chi/\ellc)\sin\theta}{\cosh(\chi/\ellc)-\cos\theta\sinh(\chi/\ellc)}\partial_\theta\,. 
\end{align*}
As emphasized in \cite{Barrow:1997vu}, it proves useful to use a different system of coordinates to recast these expressions in much
simpler forms, and we report the detailed expressions in appendix \ref{AppOpenCoordinates}.

The other possibility corresponds to model $\VIIh$. In this case the KVFs and invariant basis can be related to 
the expressions above simply as~\cite{Pontzen2010}
\beas
\gr{\xi}^{(\VIIh)}_1 &=&  \gr{\xi}^{(\V)}_1\,,\\
\gr{\xi}^{(\VIIh)}_2 &=&  \gr{\xi}^{(\V)}_2\,,\\
\gr{\xi}^{(\VIIh)}_3 &=&  \gr{\xi}^{(\V)}_3 - \ells^{-1}\partial_\phi\,,
\eeas
and
\be\label{FiveToSevenh}
\gr{e}_i^{(\VIIh)} = M_i^{\,j} \gr{e}_j^{(\V)}
\ee
with $\gr{M}$ given formally by the same expression~\eqref{spiral-matrix}, but with $z$ defined as in~\eqref{barrow-y}.

\subsubsection{Model $\IX$}\label{subsubsec:IX}

This is the only model emerging from a closed maximally symmetric space (i.e., a three-dimensional sphere) with metric
\be
\gr{g}^{\rm MS} = \dd \chi^2 + \ellc^2 \sin^2(\chi/\ellc) \dd^2 \Omega\,.
\ee
The KVFs and invariant basis are given by
\beas
\gr{\xi}_1^{(\IX)}&=& \cos \phi \sin \theta \partial_r\nonumber\\
&+&\ellc^{-1}[\cot(\chi/\ellc) \cos
\theta \cos \phi  +\sin \phi]\partial_\theta\nonumber\\
&-&\ellc^{-1}[\cot(\chi/\ellc) \csc
\theta \sin \phi  - \cos \phi \cot \theta]\partial_\phi\,,\nonumber\\
\gr{\xi}_2^{(\IX)}&=& \sin \phi \sin \theta \partial_r\nonumber\\
&+&\ellc^{-1}[\cot(\chi/\ellc) \cos
\theta \sin \phi  -\cos \phi]\partial_\theta\nonumber\\
&+&\ellc^{-1}[\cot(\chi/\ellc) \csc
\theta \cos \phi  + \sin \phi \cot \theta]\partial_\phi\,,\nonumber\\
\gr{\xi}_3^{(\IX)}&=& \cos \theta \partial_r - \ellc^{-1}[\cot(\chi/\ellc) \sin
\theta \partial_\theta + \partial_\phi]\,,\nonumber
\eeas
and by
\beas
\gr{e}_1^{(\IX)}&=& \cos \phi \sin \theta \partial_r \nonumber\\
&+&\ellc^{-1}[\cot(\chi/\ellc) \cos
\theta \cos \phi  -\sin \phi]\partial_\theta\nonumber\\
&-&\ellc^{-1}[\cot(\chi/\ellc) \csc
\theta \sin \phi  + \cos \phi \cot \theta]\partial_\phi\,,\nonumber\\
\gr{e}_2^{(\IX)}&=& \sin \phi \sin \theta \partial_r \nonumber\\
&+&\ellc^{-1}[\cot(\chi/\ellc) \cos
\theta \sin \phi  +\cos \phi]\partial_\theta\nonumber\\
&+&\ellc^{-1}[\cot(\chi/\ellc) \csc
\theta \cos \phi  - \sin \phi \cot \theta]\partial_\phi\,,\nonumber\\
\gr{e}_3^{(\IX)}&=& \cos \theta \partial_r - \ellc^{-1}[\cot(\chi/\ellc) \sin
\theta \partial_\theta - \partial_\phi]\,,\nonumber
\eeas
respectively. 

Before proceeding, note that the transformation $(\theta,\phi)\rightarrow(\pi-\theta,\pi+\phi)$ is such that
$(\gr{\xi}_i^{(\IX)},\gr{e}_i^{(\IX)})\rightarrow(-\gr{e}_i^{(\IX)},-\gr{\xi}_i^{(\IX)})$, while the metric remains invariant. Thus, the role of KVFs and invariant basis can be reversed in model $\IX$. Such inversion also happens in Bianchi~$\I$, as one can trivially check. As we will see, this has practical implications in the identification of homogeneous perturbations with spatial anisotropies in these two models.

\subsection{Linearized Bianchi equations}

We now parameterize $\gamma_{ij}$ in~\eqref{DefB-metric} as
\be\label{def-betaij}
\gamma_{ij}(t)= [e^{2 \beta(t)}]_{ij},
\ee
and we linearize equations in the time-dependent traceless matrix
$\beta_{ij}(t)$, and also in $\hat{v}_i$. When find that for the Bianchi types which admit a FLRW limit, the metric reduces to a
maximally symmetric space metric when  $\beta_{ij} =0 $ (and thus $\gamma_{ij} = \delta_{ij}$), that is we find
\be
\gr{g}^{\rm MS} = \delta_{ij} \gr{e}^i \otimes \gr{e}^j\,.
\ee
Consequently, we also recover in that case that the Ricci scalar
takes a FLRW form. This means in particular that the traceless part of the spatial
Ricci vanishes, and the value of the Ricci scalar (spatial curvature scalar) indicates to which FLRW type (open, closed or flat) it corresponds. This is reported in Table~\ref{TableType} where the values of $(A,N^1,N^2,N^3)$ are given as functions of the two length scales introduced in~\eqref{ells-def}. Neglecting sources of anisotropic stress, the background equations are
\begin{align}
\HH^2 + K & = \frac{\kappa a^2}{3} \bar \rho\,,\\
\bar \rho'+3 \HH (\bar \rho + \bar p) & = 0\,,
\end{align}
where
\be
K = -6 \hat{A}_i \hat{A}_i-\hat N^{ij} \hat  N^{ij}+\frac{1}{2}\hat{N}^{ii}\hat{N}^{jj}\,.
\ee
These are formally the same as the dynamical equations for a background FLRW metric. The curvature scale $\ellc=1/\sqrt{|K|}$ appears in the FLRW limit of the Ricci scalar whereas the spiral scale $\ells$ (whose meaning will be made clear later) does not. 

The shear is given at linear order by
\be\label{shbeta}
\hat \sigma_{ij} = \beta'_{ij}\,,
\ee
and the linear parts of the equations (in $\beta_{ij}$ and $\hat{v}_i$) are
\begin{subequations}
  \begin{align}
 \frac{1}{2}a^2\delta(\Rspatial) &= \kappa a^2 \delta \rho \,,\label{trace-eq}\\
\beta_{ij}'' + 2\HH \beta'_{ij}  & = {\cal S}_{ij}\,,\label{trace-free-eq}\\
3\hat{A}_j\beta'_{ji}+\hat{\epsilon}_{ijk}\beta'_{jl}\hat{N}_{kl} & = P_i\label{tilt-eq}\\
\delta \rho'+3\HH(\delta \rho+ \delta p) & = -(\bar \rho+\bar p)D_i \hat v^i \nonumber\\
    &= \frac{2\hat{A}^ i P_i}{\kappa a^2}\label{ConservEq-Bianchi}\\
[a^4 (\bar \rho+\bar p)\hat v_i]' &=0\label{EulerEq-Bianchi}\,.
\end{align}
\end{subequations}
The r.h.s of equations~\eqref{trace-eq} and~\eqref{trace-free-eq}
correspond to the isotropic and anisotropic contribution to the
spatial curvature due to terms linear  in $\beta_{ij}$. They are given
respectively by
\begin{align}
a^2 \delta(\Rspatial) & = 2\left(6 \hat{A}_i \hat{A}_j-2 \hat{N}^{ik} \hat{N}^{kj}  +  \hat{N}^{kk}  \hat{N}^{ij}\right)\beta_{ij}\,,\label{Iso-curvature}
\end{align}
and by
\begin{align}
{\cal S}_{ij} & = \hat{N}^{kk}\hat{N}^{\{ij\}} - 2\hat{N}^{k\{i}\hat{N}^{j\}k} - 2\hat{A}_k\hat{\epsilon}_{lk(i}\hat{N}^{j)l}\nonumber \\ 
& +2\left[2\hat{N}^{kk}\beta_{l\{i}\hat{N}^{j\}l}-\frac{1}{3}\hat{N}^{kk}\hat{N}^{ll}\beta_{ij}\right.\nonumber\\
  &+\hat{N}^{ij}\beta_{kl}\hat{N}^{kl} -   4\hat{N}^{kl}\hat{N}^{k\{i}\beta_{j\}l}\nonumber\\
  &-2\hat{N}^{ri}\hat{N}^{sj}\beta_{rs}+\frac{2}{3}\hat{N}^{kl}\hat{N}^{kl}\beta_{ij} \label{Aniso-curvature}\\
&+\left.2\hat{A}_{k}\epsilon_{kl(i}\beta_{j)r}\hat{N}^{rl} -2\hat{A}_{l}\beta_{kl}\epsilon_{kr(i}\hat{N}^{j)r}\right]\,.\nonumber
\end{align}
Repeated indices are summed, and we have introduced the notation $\{\dots\}$ for symmetric  and
trace-free tensors with respect to $\delta_{ij}$, such that all indices on the r.h.s are now manipulated with the Kronecker delta.

\begin{table*}[t]
	\centering
	\begin{tabular}{|c|c|c|c|c|c|c|}
		\hline 
		\multicolumn{2}{|c|}{$m$:} & $0$ & $+1$ & $-1$ & $+2$ & $-2$\tabularnewline
		\hline 
		\hline 
		\multicolumn{2}{|c|}{$q_{ij}^{(m)}$:} & $\frac{1}{3}\left(\begin{array}{ccc}
		1 & 0 & 0\\
		0 & 1 & 0\\
		0 & 0 & -2
		\end{array}\right)$ & $\frac{1}{\sqrt{8}}\!\left(\begin{array}{ccc}
		0 & 0 & 1\\
		0 & 0 & {\rm i}\\
		1 & {\rm i} & 0
		\end{array}\right)$ & $\frac{1}{\sqrt{8}}\!\left(\begin{array}{ccc}
		0 & 0 & -1\\
		0 & 0 & {\rm i}\\
		-1 & {\rm i} & 0
		\end{array}\right)$ & $\sqrt{\!\frac{3}{8}}\!\left(\begin{array}{ccc}
		-1 & {\rm -i} & 0\\
		-{\rm i} & 1 & 0\\
		0 & 0 & 0
		\end{array}\right)$ & $\sqrt{\!\frac{3}{8}}\!\left(\begin{array}{ccc}
		-1 & {\rm i} & 0\\
		{\rm i} & 1 & 0\\
		0 & 0 & 0
		\end{array}\right)$\tabularnewline
		\hline 
		\multirow{3}{*}{\parbox{0.05\linewidth}{\vspace{1cm} $\VIIo$}} & $\ell_{s}^{2}{\cal S}^{(m)}$ & $0$ & $0$ & $0$ & $-4$ & $-4$\tabularnewline
		& $\ell_{s}^{2}{\cal R}^{(m)}$ & $0$ & $0$ & $0$ & $0$ & $0$\tabularnewline
		& $\sqrt{8}\ell_{s}{\cal P}_{i}^{(m)}$ & $\left(\begin{array}{c}
		0\\
		0\\
		0
		\end{array}\right)$ & $\left(\begin{array}{c}
		-{\rm i}\\
		1\\
		0
		\end{array}\right)$ & $\left(\begin{array}{c}
		-{\rm i}\\
		-{\rm 1}\\
		0
		\end{array}\right)$ & $\left(\begin{array}{c}
		0\\
		0\\
		0
		\end{array}\right)$ & $\left(\begin{array}{c}
		0\\
		0\\
		0
		\end{array}\right)$\tabularnewline
		\hline 
		\multirow{4}{*}{\parbox{0.05\linewidth}{\vspace{1cm} $\V$}} 
		& $\ell_{c}^{2}{\cal S}^{(m)}$ & $0$ & $0$ & $0$ & $0$ & $0$\tabularnewline
		& $\ell_{c}^{2}{\cal R}^{(m)}$ & $-8$ & $0$ & $0$ & $0$ & $0$\tabularnewline
		& $\sqrt{8}\ell_{c}{\cal P}_{i}^{(m)}$ & $\left(\begin{array}{c}
		0\\
		0\\
		-4\sqrt{2}
		\end{array}\right)$ & $\left(\begin{array}{c}
		3\\
		3{\rm i}\\
		0
		\end{array}\right)$ & $\left(\begin{array}{c}
		-3\\
		3{\rm i}\\
		0
		\end{array}\right)$ & $\left(\begin{array}{c}
		0\\
		0\\
		0
		\end{array}\right)$ & $\left(\begin{array}{c}
		0\\
		0\\
		0
		\end{array}\right)$\tabularnewline
		\hline 
		\multirow{4}{*}{\parbox{0.05\linewidth}{\vspace{1cm} $\VIIh$}} & $\ell_{c}^{2}{\cal S}^{(m)}$ & $0$ & $0$ & $0$ & $-4/h-4{\rm i}/\sqrt{h}$ & $-4/h+4{\rm i}/\sqrt{h}$\tabularnewline
		& $\ell_{c}^{2}{\cal R}^{(m)}$ & $-8$ & $0$ & $0$ & $0$ & $0$\tabularnewline
		& $\sqrt{8}\ell_{c}{\cal P}_{i}^{(m)}$ & $\left(\begin{array}{c}
		0\\
		0\\
		-4\sqrt{2}
		\end{array}\right)$ & $\left(\begin{array}{c}
		3-{\rm i}/\sqrt{h}\\
		3{\rm i}+1/\sqrt{h}\\
		0
		\end{array}\right)$ & $\left(\begin{array}{c}
		-3-{\rm i}/\sqrt{h}\\
		3{\rm i}-1/\sqrt{h}\\
		0
		\end{array}\right)$ & $\left(\begin{array}{c}
		0\\
		0\\
		0
		\end{array}\right)$ & $\left(\begin{array}{c}
		0\\
		0\\
		0
		\end{array}\right)$\tabularnewline
		\hline 
		\multirow{2}{*}{$\I$} & ${\cal S}^{(m)}$ & $0$ & $0$ & $0$ & $0$ & $0$\tabularnewline
		& ${\cal R}^{(m)}$ & $0$ & $0$ & $0$ & $0$ & $0$\tabularnewline
		\hline 
		\multirow{2}{*}{$\IX$} & $\ell_{c}^{2}{\cal S}^{(m)}$ & $-8$ & $-8$ & $-8$ & $-8$ & $-8$\tabularnewline
		& $\ell_{c}^{2}{\cal R}^{(m)}$ & $0$ & $0$ & $0$ & $0$ & $0$\tabularnewline
		\hline 
	\end{tabular}
	\caption{Bianchi $svt$ modes and the quantities ${\cal R}^{(m)}$, ${\cal S}^{(m)}$,
	and ${\cal P}_{i}^{(m)}$ for all Bianchi models considered in
        this work. For models I and IX, ${\cal P}_{i}^{(m)}=0$, and
        thus not shown. This reproduces Table {\rm II} of PC10 up to variations of conventions in definitions.}\label{svtTable}
\end{table*}

\subsection{Homogeneous $svt$ modes}\label{subsec:svt}

In order to identify $\beta_{ij}$ with homogeneous metric perturbations, we proceed by decomposing the former in a similar fashion to the decomposition of $\delta g_{ij}$. Just as the Scalar, Vector and Tensor (SVT) modes appearing in~\eqref{SVT-metric} are defined with respect to their transformation properties under
rotations around $\gr{k}$ -- the wave-vector of the perturbation -- we
can introduce scalar, vector and tensor modes of the shear (henceforth
$svt$ modes) with respect to their transformations under rotations
around some  direction $\gr{e}_i$ of the invariant
basis~\cite{Pontzen2010}. For models $\I$ and $\IX$ it does not matter
which direction we choose, since any rotation will preserve
${C^k}_{ij}$ in these models. On the other hand, the constants of
structure in models $\V$, $\VIIo$ and $\VIIh$ have a residual symmetry
given by rotations around the vector $\gr{A}\propto\gr{e}_3$. Since
the construction in this section is general to all Bianchi models, we
shall omit a subscript in the co-vectors $e_i$ to designate the model
they belong to, so as to alleviate the notation.
We thus choose the latter as a fiducial direction and introduce a complex basis
\be
e^{(\pm)}_i(\gr{e}_3) \equiv \frac{(e^1_i\mp \ii e^2_i)}{\sqrt{2}}\,.
\ee
in terms of which we introduce the following tensor polarization basis
\begin{align}
\begin{split}\label{Defqij}
q^{(0)}_{ij}(\gr{e}_3) & \equiv (-e^3_ie^3_j+\delta_{ij}/3)\,, \\
q^{(\pm 1)}_{ij}(\gr{e}_3) & \equiv \pm e^3_{(i}e^{(\mp)}_{j)}\,,\\
q^{(\pm 2)}_{ij}(\gr{e}_3) & \equiv -\sqrt{\frac{3}{2}}e^{(\mp)}_ie^{(\mp)}_j\,.
\end{split}
\end{align}
In what follows, we will omit the dependence on $\gr{e}_3$ whenever there is no chance of confusion. Note also that, because $\gr{e}_3$ and $\gr{e}^\pm$ form a general triad frame, $q^{(m)}_{ij}$ will in general depend on the spacetime point, whose explicit dependence we shall also omit.

The above polarization tensors allow us to write the shear as
\be\label{betam}
\beta_{ij} = \sum_{m=-2}^2 \beta_{(m)}q^{(m)}_{ij}
\ee
with the values of $m$ corresponding to scalar ($m=0$), vector ($m=\pm1$) and tensor ($m=\pm2$) modes. Note that all modes are constructed such that
\be\label{qminusm}
q^{(-m)}_{ij} =(-1)^m q^{(m)\star}_{ij} \,,
\ee
and hence the reality of $\beta_{ij}$ implies
\be\label{betasym}
\beta_{(-m)} = (-1)^m \beta^{\star}_{(m)}\,.
\ee
This means that, in practice, we only need to consider physical effects for the case $m>0$. 

Because the Bianchi models we are considering all have a maximally symmetric 3-space as their isotropic limits, the first line of equation~\eqref{Aniso-curvature} vanishes for all models of Table~\ref{TableType}, as one can easily check. We can thus write
\be\label{DecSij}
{\cal S}_{ij} = \sum_{m=-2}^2 {\cal S}^{(m)} \beta_{(m)}q^{(m)}_{ij}\,.
\ee
The isotropic spatial curvature, on the other hand, splits into a background plus perturbation as
\be
a^2\Rspatial = a^2\Rspatial_{\rm FLRW} + \delta(a^2\Rspatial)\,,
\ee
where 
\be\label{Rspatial-m}
\delta(a^2\Rspatial) = \sum_{m=-2}^2{\cal R}^{(m)}\beta_{(m)}\,,
\ee
and
\begin{align}
{\cal R}^{(m)} & = {\cal R}^{ij}q^{(m)}_{ij}\,,\\
{\cal R}_{ij}  & = 2\left(6 \hat{A}_i \hat{A}_j-2 \hat{N}^{ik} \hat{N}^{kj}  +  \hat{N}^{kk} \hat{N}^{ij}\right)\,.\nonumber
\end{align}
Likewise, the tilt $P_i$ can be decomposed as
\be\label{Pi-svt}
P_i = \sum_{m=-2}^2 \beta'_{(m)}{\cal P}^{(m)}_i\,,
\ee
with
\be
{\cal P}^{(m)}_i = 3\hat{A}_jq^{(m)}_{ji}+\hat{\epsilon}_{ijk}q^{(m)}_{jl}\hat{N}_{kl}\,.
\ee
The $svt$ modes for the tensors ${\cal S}_{ij}$, ${\cal R}_{ij}$ and ${\cal P}_i$ for all Bianchi models are summarized in Table~\ref{svtTable}.

\section{Bianchi and FLRW modes related}\label{SecMatch}

\subsection{Matching the perturbations}

\cpsilent{We are now ready to find the correspondence between Bianchi degrees of
freedom and FLRW metric perturbations. In other words, we can now identify the modes in the expansion \eqref{ReplacementRuleBardeen} which we must 
consider in \eqref{SVT-metric} so as to obtain the matching
\be\label{CoreMatching}
\delta g_{ij}\to 2a^2\beta_{ij}\,.
\ee
Since $\beta_{ij}$ is traceless, one can already set $\phi \to 0$. On the Bianchi side, we have seen that the shear can be decomposed 
as in Eq.~\eqref{betam}. As we detail in Appendix~\ref{Appzeta}, we show that
\be\label{qQplus}
q^{(m)}_{ij}(\chi,\gr{n}) = \frac{\mygamma_2}{\mygamma_m}\, Q^{(2m)}_{ij}(\chi,\gr{n};\nu_m,\myzeta_\ell^m),
\ee
where the constants $\mygamma_m$ are defined as (see~\cite{Pitrou:2019ifq} for details)
\be\label{DefGamma}
\mygamma_m \equiv \prod_{i=1}^m \frac{k \ellc}{\sqrt{(\nu_m \ellc)^2-\barK i^2}}\,,
\ee
with the understanding that $\mygamma_0=1$. Defining
\be\label{Htobeta}
H_T^{(m)} \equiv \beta_{(m)}\frac{\mygamma_2}{\mygamma_m} \,.
\ee
It implies that the identification \eqref{CoreMatching} is made with the discrete sum
\be\label{CoreMatching2}
\frac{\delta g_{ij}}{2 a^2} = \sum_{m=-2}^2 H_T^{(m)}(\nu_m,\eta)Q^{(2m)}_{ij}(\nu_m,\myzeta_\ell^m)\,.
\ee
Let us stress already that the identification of model $\IX$ is a special case since there are
only tensor modes. That is, the Bianchi modes $q_{ij}^{(m)}$ with
$m=0,1,-1$ do not map to scalar and vector harmonics, but instead to
sums of tensor harmonics. We shall treat this case separately.

We still need to specify with which modes $\nu_m$ and with which set of constants
$\myzeta_\ell^m$ the matching \eqref{qQplus} holds.  The matching of the positive and negative $m$ are necessarily related
since negative values can be obtained by the reality
condition~\eqref{qminusm}. Using Eq. (6.8) of \cite{Pitrou:2019ifq} for the
complex conjugation of an harmonic, we can check that we obtain
\be
\left[ Q_{ij}^{(2m)}(\nu_m,\myzeta_\ell^{m})\right]^\star = (-1)^m Q_{ij}^{(2,-m)}(\nu_{-m},\myzeta_\ell^{-m})\nonumber
\ee
provided that the conditions
\be\label{Constraintsnuzeta}
\nu_{-m}= -\nu_m^\star\,,\qquad \myzeta_\ell^{-m} = (-1)^\ell \myzeta_\ell^{m\star}
\ee
are satisfied.

In order to find for each $m$ the mode $\nu_m$ and the set of coefficients $\myzeta_\ell^m$ which
define the pseudo plane-wave, we first determined the $\nu_m$ by comparing equations~\eqref{trace-free-eq}
and~\eqref{tilt-eq} to the linearized Einstein equations given in
\S\ref{sec:lpt}, and asking the constraints \eqref{Constraintsnuzeta}
to be satisfied. The shear evolution in Bianchi models maps
to either tensor, vector or traceless scalar perturbations in
synchronous gauge and we report details of the matching in appendix~\ref{MatchEquations}.
For the flat and open cases (corresponding to types
$\I,\V,\VIIo,\VIIh$), the modes must be
\be\label{Valuenum}
\nu_m = \frac{m}{\ells} + \frac{\ii}{\ellc}
\ee
with $k_m = \nu_m$ in the $\I$ and $\VIIo$ cases. For the closed case
(Bianchi $\IX$) we obtain
\be
\nu_{\pm 2} = \pm \frac{3}{\ellc}\,.
\ee
Given that the $\nu_m$ can be complex in the open case, it is
understood that the radial functions are obtained through analytic
continuation corresponding to the presence of a super-curvature modes
(see \S\ref{SuperCurvature}). Technically, this implies that the
electric and magnetic radial functions are no more real valued.

We must also specify the coefficients $\myzeta_\ell^m$ since some models are matched
with pseudo plane-waves. In Appendix~\ref{Appzeta} we detail how
these constants are found for models $\VIIo$, $\V$ and $\VIIh$, and
they are (up to a global constant factor)
\beas\label{myzeta}
\myzeta_{m}^{\pm m} &=& (\pm 1)^m\slabel{myzetaa}\\
\myzeta_{\ell}^{\pm m}&=&\myzeta_{\ell-1}^{\pm
  m}(-\ii)\frac{\sqrt{\ell^2 + (\nu_{\pm m} \ellc)^2}}{(\ell+1) \mp m \ii/\sqrt{h}}\,.\slabel{myzetab}
\eeas
Using the identity~\eqref{l2nu2}, it can be recast in the form
\be
\myzeta_\ell^{\pm m} = (\pm 1)^m \prod_{p=m+1}^\ell (-\ii)\sqrt{\frac{(p-1)\pm m \ii/\sqrt{h}}{(p+1)\mp m \ii/\sqrt{h}}}\,.
\ee
The case  $m=0$ is special since in \eqref{myzeta} one finds
\be\label{zerocase}
\myzeta_0^0 =1,\quad \myzeta_{\ell\geq 1}^0=  (-\ii)^\ell \frac{\sqrt{1+(\nu \ellc)^2}}{\sqrt{2
    \ell (\ell+1 )}}\,.
\ee
Note that there is a factor $\sqrt{1+(\nu_0 \ellc)^2}=0$ in all ratios $\myzeta_\ell^0/\myzeta_0^0$. One could think that this is problematic
since the ratios $\myzeta_\ell^m/\myzeta_j^m$ enter the very definition of a pseudo plane-wave \eqref{DefQfromsuml}. For instance, in
$Q^{(20)}_{ij}$ one would encounter $\myzeta_0^0/\myzeta_2^0 \sim 1/\sqrt{1+(\nu_0 \ellc)^2}$. However the factor $\sqrt{1+(\nu_0 \ellc)^2}$ is nothing but $k$ given the relation
\eqref{DefNu}, and it appears that there is a compensating divergence
in the radial functions. More precisely it can be checked that ${}_s\alpha_0^{(jm)} \sim \sqrt{1+(\nu_0
  \ellc)^2}$ for $k \to 0$ (except ${}_0 \alpha_0^{(00)} \to 1$). In
practice, one must keep track of all factors of $k$ and take the limit $k \to 0$ at the end. More rigorously, one could have redefined radial functions to be
\be
\frac{\myzeta_\ell^m}{\myzeta_j^m} {}_s \alpha_\ell^{(jm)}
\ee
as these quantities would never possess any (apparent) divergences.

The results for models $\VIIo$ and $\V$ can be obtained by letting $\ellc\rightarrow\infty$ (or $h\rightarrow 0$) and $\ells\rightarrow\infty$ (or $h\rightarrow
\infty$), respectively. For model $\VIIo$ this reduces to
$\myzeta_\ell^{\pm |m|} = (\pm 1)^\ell$. In the Bianchi $\V$ case, we
also find \eqref{zerocase} when $m=0$, and when $m \neq 0$ we get
\beas
\zeta_\ell^{\pm m}=(\pm 1)^m (-\ii)^{\ell-m}\sqrt{\frac{m(m+1)}{\ell(\ell+1)}}\,.
\eeas
As for models $\I$ and $\IX$, we find $\myzeta^{\pm m}_\ell=\delta_\ell^2$ 
that is the sum is reduced to the lowest term with  $\ell=2$. In the case of model $\I$, one could alternatively consider that $\myzeta^{\pm m}_\ell=(\pm 1)^\ell$ since for $\ell \geq 3$ all
radial functions vanish for $k_m=0$. The pseudo plane-waves harmonics
needed for the matching in the various models are summarized in table \ref{TableMatching}.

Let us comment that the scalar mode ($m=0$) is special. Indeed we
found that the $\nu_0$ and $\myzeta_\ell^0$ are the same for models
$\V$ and $\VIIh$, and also  the same  for  models $\I$ and $\VIIo$. This is
because there is a spiraling structure in models $\VIIo$ and $\VIIh$
(with typical spiral scale $2\pi\ells/|m|$) which is absent for the scalar mode ($m=0$). 

Note that the prefactor $\mygamma_2/\mygamma_m$ is always compensated by an
opposite factor $\mygamma_m/\mygamma_2$ in the ${}_s g^{(2m)}$ which
enter in the definition of harmonics from radial functions
\eqref{Qljm}. This was expected since this prefactor was precisely added to remove any
divergence so as to reach constant matrices for the $q_{ij}^{(m)}$.

\begin{center}
	\begin{table}
		\begin{centering}
			\begin{tabular}{|c|c|c|}
				\hline 
				& $\nu$ or $k$ & $\myzeta_{\ell}^{\pm m}$\tabularnewline
				\hline 
				\hline 
				$\I$ & $k_m\rightarrow0$ & $\delta_\ell^2$\tabularnewline
				\hline 
				$\VIIo$ & $k_m=\frac{m}{\ell_{s}}$&   $(\pm 1 )^\ell$\tabularnewline
				\hline 
				$\IX$ & $\nu_{\pm 2}=\pm\frac{
                                        3}{\ell_{c}}\,$ &
                                                          $\delta_\ell^2$,
                                                          $|m|=2$ only \tabularnewline
				\hline 
				$\V$ &
                                       $\nu_m=\frac{\text{i}}{\ell_{c}}$
                          &$(\pm 1)^m (-\ii)^{\ell-m}\sqrt{\frac{m(m+1)}{\ell(\ell+1)}}$\tabularnewline
				\hline 
				$\VIIh$ &
                                          $\nu_m=\frac{m}{\ell_{s}}+\frac{\text{i}}{\ell_{c}}$
                          & $(\pm)^m\prod_{p=m+1}^{\ell}(-\ii)\sqrt{\frac{(p-1)\sqrt{h}\pm m{\rm i}}{(p+1)\sqrt{h}\mp m{\rm i}}}$\tabularnewline
				\hline 
			\end{tabular}
			\par\end{centering}
		\caption{Summary of modes ($k_m$ or $\nu_m$) and
                  pseudo plane-wave constants ($\zeta_\ell^m$) found
                  in this section. It is understood that the
                  $\myzeta_\ell^0$ for the $\V$ and $\VIIh$ models are
                \eqref{zerocase}.}\label{TableMatching}
\end{table}
\par\end{center}

\subsection{Transformation properties}\label{SecTransfo}

When writing the constants of structure in canonical form in Table~\ref{TableType}, we have
used the possibility of a global rotation and of parity inversion. Hence, once we have identified the harmonics corresponding
to the $s,v,t$ modes, we must explore the effect of global rotations and parity inversion to exhaust all possible FLRW metric perturbations
matching Bianchi models.

In Table I of \cite{Pitrou:2019ifq} are gathered the transformation rules
of the harmonics ${}^\ell Q^{(jm)}_{I_j}(\nu,\eta_\ell^m)$ (defined with respect to the zenith axis) under inversion of a single individual axis ($x$, $y$ or
$z$) axis. For the pseudo plane-waves used in the matching (with
$j=2$), the transformation rules can be deduced (see Section 6.4 of
\cite{Pitrou:2019ifq}) and we summarize them in table~\ref{TableParity} for
clarity of the following discussion. 
\begin{table}[!htb]
\begin{tabular}{|c|c|c|c|}
\hline
& $x \to -x$ & $y \to -y$ & $z \to -z$\\
\hline
  Factor $(-1)^m$&&yes&yes\\
  $\zeta_\ell^m \to (-1)^\ell \zeta_\ell^m $&&&yes\\
   $Q_{I_j}^{(jm)}\to Q_{I_j}^{(j,-m)}$&yes&yes&\\
$Q_{I_j}^{(jm)}(\nu)\to Q_{I_j}^{(jm)}(-\nu) $&&&yes\\
  \hline
 \end{tabular}
\caption{Transformation rules for harmonics (defined with the zenith direction) under the inversion of a
  single axis. \label{TableParity}}
\end{table}

\subsubsection{Isotropic constants of structure}

When the constants of structure of a given model are invariant under
arbitrary rotations (such as in models $\I$ and $\IX$), any mode $s$,
$v$ or $t$ (or, equivalently, any basis $q^{(m)}_{ij}$ for $|m|=0$,
$1$ or $2$) can be constructed from any other by appropriate linear combinations of rotated modes. One can easily verify that\be\label{rotated-q0}
q^{(0)}_{ij}(\gr{e}_3)  \propto  \sum_{m=\pm 2}q^{(m)}_{ij}(\gr{e}_1) + q^{(m)}_{ij}(\gr{e}_2)\,,
\ee
where we use $ e^{(\pm)}_i(\gr{e}_1) \equiv (-e^3_i\mp \ii e^2_i)/\sqrt{2}$ and $ e^{(\pm)}_i(\gr{e}_2) \equiv (-e^3_i\pm \ii e^1_i)/\sqrt{2}$.
Similar combinations can be used to produce the $q^{(\pm1)}_{ij}(\gr{e}_3)$. As a consequence all modes have the same
dynamics and this is checked in Table~\ref{svtTable} where the ${\cal
  S}^{(m)}$ are the same for all $m$ in models $\I$ and $\IX$. Thus, the anisotropies of every Bianchi $\I$ model can be seen as combinations of homogeneous gravitational
waves, which are exhausted by the fives parameters $\beta_{(m)}$ which, given \eqref{betasym}, are fives degrees of freedom of the
model. Any global rotation leads only to a transformation of the $\beta_{(m)}$ by a constant phase.

The freedom in the point of view about the nature of the modes results from the non-uniqueness in the definition of SVT modes when perturbations do
not decay at infinity~\cite{Stewart:1990fm}. However, since Bianchi $\IX$ arises as perturbations of a closed FLRW universe, which has compact spatial sections, 
there is no such freedom of interpretation and the $s$, $v$ modes {\it must} be
considered as sums of tensors modes. In other words, they are obtained from sums of rotations of $q^{(\pm 2)}_{ij}(\nu_{\pm 2})$.

In practice, when exploring the implications of Bianchi models on
observables, one might choose to lose complete generality and to
focus on the effect of a single gravitational wave. Hence, one starts from the tensor mode aligned with the zenith direction
$\beta_{(\pm 2)}{}^{\ell=2}Q_{ij}^{(2,\pm 2)}(\nu_{\pm 2})$ and explores the range of Euler
angles for the rotation of the plane-wave axis. When doing so, one does not need to explore the freedom of the $\gamma$ Euler angle, (i.e., of
rotations of the wave around its wave vector $\gr{\nu}$) since this is degenerate with the phases in the $\beta_{(m)}$.

\paragraph{Bianchi $\I$.}

In the simple $\I$ models, one has $\nu_m = 0$ and only $\ell=2$, so we infer
from Table \ref{TableParity} that the tensor mode with zenith direction is invariant under inversion of the
$z$ axis. Also since $\nu_m=\nu_{-m}$ we deduce that an inversion of
either the $x$ or the $y$ axis interchanges the $m$ and $-m$
contributions, leaving the whole metric perturbation invariant. Hence
the $s,v,t$ modes are invariant under inversion of any axis and in
particular of global parity.

\paragraph{Bianchi $\IX$.}

The case of Bianchi $\IX$ is different. Indeed, since $\nu_{-2} =
-\nu_{2}$ one loses the parity invariance as seen on Table \ref{TableParity}. The tensor mode with zenith
axis is only invariant by simultaneous inversion of two axes among
$x,y,z$. Physically, this happens because such tensor mode in Bianchi
$\IX$ is a standing circularly polarized wave~\cite{King:1991jd}; intuitively this is like a spiraling
structure in the zenith direction (and which has a spiraling effect on
observables~\cite{Pontzen2009}), and any rotation which flips this spiraling
direction leaves the system invariant. As found in
\cite{King:1991jd} and summarized in Appendix~\ref{App9}, the difference between the type $\IX$ and type $\I$  is related to the
fact that the canonical choice of constants of structure selects only
one type of chirality for the circularly polarized wave (one could
invert all signs of the constants of structure). The Bianchi $\IX$ is a sum of (rotated) tensor harmonics of the same
chirality~\cite{King:1991jd}. To exhaust all possible perturbations of
the type $\IX$ one then needs to consider a global parity
transformation which changes the chirality.

\begin{figure*}[!htb]
\includegraphics[scale=0.5]{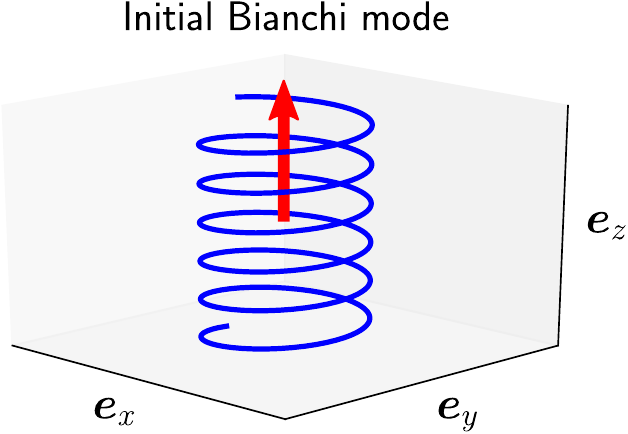}\; \includegraphics[scale=0.5]{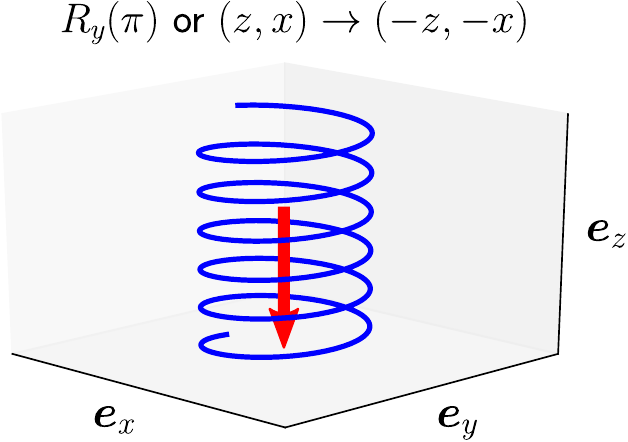}\; \includegraphics[scale=0.5]{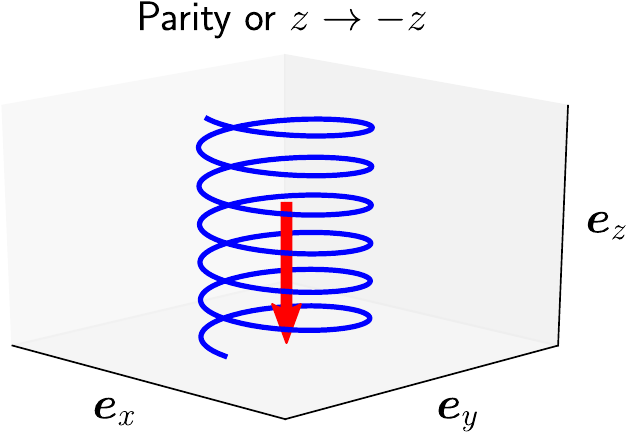}
\caption{Schematic representation of the transformation properties of
  the spiraling and focusing structures, under inversions of some
  axes. The type $\VIIo$ has only a spiraling structure (represented by the blue screw thread), whereas the
  type $\V$ has only a focusing effect (represented by the red arrow). Type $\VIIh$ has both features.}\label{FigScrew1}
\end{figure*}

\subsubsection{Anisotropic constants of structure}

For anisotropic constants of structure, that is, for models $\V,\VIIo,\VIIh$, one has the five degrees of
freedom of the amplitude $\beta_{(m)}$, on top of which we must also allow for a
general direction of the special axis used when writing the constants in canonical form. This brings two other degrees of freedom (the
angles of the special axis direction) since any rotation around the special axis is degenerate with phases in the
$\beta_{(m)}$. It is then instructive to look at the properties under inversion of
the axis, since it highlights the effects that modes have on observables. As we now detail, a global parity inversion is needed to
explore the whole range of $\VIIo$ and $\VIIh$ types but this is not needed for the $\V$ type.

\paragraph{Bianchi $\V$.}

The associated modes have the property $\nu_{-m} = \nu_m$ and $\zeta_\ell^{-m} = \zeta_\ell^m$. Hence from Table \ref{TableParity}  we find that it is invariant under inversion of either the $x$ or the $y$ axis. However it is not invariant under
inversion of the $z$ axis. This is because Bianchi $\V$ has a
focusing effect on observables in the special $z$ direction~\cite{Barrow:1997vu,Pontzen2009}.  This
contrasts with the cases of Bianchi $\VIIo$ and $\VIIh$. Since a
rotation of angle $\pi$ around e.g. the $x$ axis inverts the $y$ and
$z$ axis, one deduces that global parity inversion does not bring new
models when exploring all possible directions.

\paragraph{Bianchi $\VIIo$.}

We still have the property $\zeta_\ell^{-m} = (-1)^\ell\zeta_\ell^m$ but we now have ${\nu_{-m} = -\nu_m}$. So from Table \ref{TableParity}  we find that the
perturbation is invariant under the joint inversion of any two of the three $x,y,z$ axes. This is the same symmetry as the one already found
in the $\IX$ case. In fact, the Bianchi $\VIIo$ modes with $m \neq 0$ are also equivalent to a standing circularly polarized standing wave, whose wavelength is free
and controlled by $\ells$. And as for the $\IX$ case, a global parity inversion, that is a change of sign in the constants of structure,
inverts the chirality of the standing wave. 

\paragraph{Bianchi $\VIIh$.}

The $\VIIh$ models possess both the properties of models $\V$ (that is
a focusing direction) and $\VIIo$ (a spiraling structure from a circularly polarized standing wave in the $m \neq 0$ modes).
An inversion of the $z$ axis inverts the focusing direction but not the
spiraling structure. An inversion of either the $x$ or $y$ axis
inverts the chirality of the spiraling structure but leaves the focusing
direction unchanged. Hence a global parity inversion (a change of sign in the constants of structure), combined with a
rotation of angle $\pi$ around the $x$ axis, inverts the chirality of
the standing wave while keeping the same focusing direction.
The transformation properties of modes in Bianchi $\VIIh$ are illustrated in Fig.~\ref{FigScrew1}.

\section{Cosmological implications}\label{Sec-Implications}

We now illustrate the power of the identification to
compute theoretical predictions for observables in various Bianchi
models. Indeed, up to now the effect of a Bianchi space-time on the CMB was
computed with independent codes from the ones used for the usual
stochastic perturbations around a FLRW background \cite{Saadeh:2016bmp,Saadeh:2016sak}}. With the identification~\eqref{CoreMatching} of Bianchi modes as
FLRW perturbations in synchronous gauge, it is possible to compute the linear transfer functions from
initial conditions to observable multipoles in the same framework. The modes identified as a Bianchi perturbations shall not correlate statistically with other FLRW perturbations since the one-point
average of usual fluctuations vanishes. However it should contribute in the three point function since the two-point function average of
usual fluctuations are related to the non-vanishing power spectrum. In
the next section we review the dynamics of a Bianchi perturbations and
recall that only some modes are regular~\cite{Pontzen2009} and thus credible as a possible
large scale anisotropy. Then we detail the computation of the CMB
multipoles and show that it requires only to adapt the usual Boltzmann
hierarchy to take into account the fact that we have pseudo plane-waves instead of usual plane-waves.

\subsection{Bianchi perturbation dynamics}

Since our method is a based on a small shear approximation, it is important to discuss the existence of solutions to Eq.~\eqref{trace-free-eq} which are finite at high redshifts. Such solutions were extensively discussed in Refs.~\cite{Pontzen2009,Pontzen2010}, and so we will just give a brief summary. The main equation is
\be\label{betadynamic}
\beta''_{(m)}+2{\cal H}\beta'_{(m)}-{\cal S}^{(m)}\beta_{(m)}=0
\ee
which follows from \eqref{trace-free-eq} with \eqref{betam} and
\eqref{DecSij}.

Being a homogeneous second order differential equation, all solutions are linear combinations of two solutions. In the case 
${\cal S}^{(m)}=0$, one solution is a pure constant with no observable
effect, while the other is diverging at early times. Hence, these modes are considered as being
irregular~\cite{Pontzen2009}, and they are usually rejected on the basis that they
would not pass observational tests at early times. Furthermore, it is also difficult to find a
natural mechanism to generate their initial conditions. 

In the $\VIIo$ and $\VIIh$ models, we have ${\cal S}^{(\pm 2)} \neq
0$, hence these modes have a different dynamics. In a matter dominated
era, one has $a \propto \eta^2$ and thus $2 \HH = 4/\eta$. The regular solution of \eqref{betadynamic} for the tensor modes in these
models is then
\be\label{BetaSol}
\beta_{(\pm 2)} \propto \frac{j_1(\omega_\pm \eta)}{\omega_\pm \eta}\,,\quad
\omega_\pm \equiv \frac{2}{\ells}\sqrt{1\pm \ii (\ells/\ellc)}\,.
\ee 
Given that $\omega_\pm$ is complex valued, this solution has to be
understood from the analytic continuation of the spherical Bessel function
$j_1$. When $\ells \gg \ellc$, the solution is regular as long as $\eta \ll
\sqrt{\ellc \ells}$, but in that case it is a constant solution and we recover
the dynamics of a Bianchi $\V$. In the opposite regime where $\ells \ll
\ellc$, the solution is regular for $\eta \ll \ellc$ (corresponding to
the isotropic curvature scale remaining super-horizon) and the
solution results in damped oscillations. Type $\IX$ also has a regular solution with damped oscillations, since ${\cal S}^{(\pm 2)}
=-8/\ellc^2$ and one finds that \eqref{BetaSol} holds with $\omega_\pm = \sqrt{8}/\ellc$ in a matter dominated era.

Given the matching with FLRW perturbations, we can interpret these regular solutions as frozen tensor modes which become dynamical when their
wavelength becomes sub-horizon. Note that the non regular modes could still be interesting for cosmological models with a homogeneous (or very large scale)
anisotropic stress which would then source the right-hand side of
\eqref{betadynamic}, since in this case one could set vanishing initial conditions and
still have observational signatures (see, e.g,~\cite{Pereira:2015jya}).

\subsection{CMB}
\label{SecCMB}

\subsubsection{Boltzmann hierarchy}\label{SecBoltzmann2}

We define CMB multipoles exactly like in section 7.2 of \cite{Pitrou:2019ifq} (which is also the notation used in~\cite{TAM1,TAM2}). Moreover,
we use the notation of section 7.3 of \cite{Pitrou:2019ifq} for gravitational and collisional terms. When using pseudo plane-waves (that is with $\myzeta_\ell^m \neq {\rm
  const.}$), as is the case for the harmonics which realize the matching with the Bianchi models, the Boltzmann hierarchy [Eqs. (7.30) of \cite{Pitrou:2019ifq}], is modified by the rules
\begin{align}
\begin{split}\label{Modifiedkappa}
{}_s \kappa_j^m &\to  {}_s \kappa_j^m
\frac{\myzeta_{j}^m}{\myzeta_{j-1}^m}\,,\\
{}_s \kappa_{j+1}^m &\to  {}_s \kappa_{j+1}^m \frac{\myzeta_{j}^m}{\myzeta_{j+1}^m}\,.
\end{split}
\end{align}
It is also implied that the $\nu$ which appears in the definition of
${}_s \kappa_\ell^m$ [Eq.~(3.19) of \cite{Pitrou:2019ifq}]  is $\nu_m$. The
modification \eqref{Modifiedkappa} arises because, for pseudo plane-waves, the recursive
relations for the normal modes, e.g. Eq. (7.29) of \cite{Pitrou:2019ifq}, are now satisfied for the combination ${}_s \overline{G}^{(jm)} \myzeta_j^m$. Consequently, we find that the integral solutions [Eqs. (7.33) of
\cite{Pitrou:2019ifq}] are modified by an extra factor
$\myzeta_{j}^m/\myzeta_{j'}^m$ on their right hand sides, that is
\bea
\frac{\Theta_j^m(\eta_0)}{2j+1} &=&  \int_0^{\eta_0}
\dd \eta {\rm e}^{-\tau} \sum_{j'} \left({}^\Theta{\cal C}_{j'}^m  + {\cal
    G}_{j'}^m \right)\nonumber\\
&&\qquad\times\frac{\myzeta_{j}^m }{\myzeta_{j'}^m}\,{}_0 \bar \epsilon_j^{(j'm)}(\chi)\,,\label{LOS}\\
\frac{ E_j^m(\eta_0)}{2j+1} &=&  \int_0^{\eta_0} \dd \eta {\rm e}^{-\tau} \sum_{j'} {}^E{\cal  C}_{j'}^m\,\frac{\myzeta_{j}^m}{\myzeta_{j'}^m}\, {}_2 \bar \epsilon_j^{(j'm)}(\chi)\,,\nonumber\\
\frac{B_j^m(\eta_0)}{2j+1} &=&  \int_0^{\eta_0} \dd \eta {\rm e}^{-\tau} \sum_{j'} {}^E{\cal C}_{j'}^m\,\frac{ \myzeta_{j}^m }{\myzeta_{j'}^m}\,{}_2 \bar \beta_j^{(j'm)}(\chi)\,,\nonumber
\eea
with $\chi = \eta_0-\eta$.

The line of sight integral solution \eqref{LOS} is only formal because the sources depend on the multipoles
themselves and to compute them one needs to rely on the
hierarchy. However in practice it is sufficient to solve the Boltzmann
hierarchy for a small number of multipoles, since the sources are
restricted to multipoles with $j \leq 2$, and to then use the integral solutions for all multipoles.

The case of tensor modes ($m=2$), which are also the only regular modes,
is interesting. The temperature quadrupole is sourced by an integrated
Sachs-Wolfe effect (ISW) from tensor modes, and is damped by
collision. When Compton scattering is inefficient, the radial
functions account for the effect of free streaming, and this reveals
the focusing and spiraling patterns of the Bianchi modes if present in
the perturbation. Scattering of the temperature quadrupole also generates
the electric quadrupole of polarization, and subsequent free-streaming feeds higher multipoles of both the electric and
magnetic types.

We notice on the structure of the Boltzmann hierarchy that the scalar
mode ($m=0$) of the Bianchi $\V$ and $\VIIh$ cases is
special. Indeed, in that case ${}_s \kappa_1^0 \myzeta_1^0/\myzeta_0^0 = 0$, which implies that the temperature
  monopole does  not feed the temperature dipole. This is expected since a homogeneous monopole
  cannot have a divergence. However, there is a homogeneous divergence
  of radiation velocity which feeds the monopole. This is because these Bianchi models are tilted and the fluid velocity (proportional
  to ${\cal P}_i^{(0)}$) is not normal to the foliation of the hypersurfaces.

\subsubsection{Comparison with \cite{Pontzen2007}}

Our results can be compared with those of \cite{Pontzen2007} using a different method. From
\eqref{myzetab} (or \ref{zerocase} when $m=0)$ and the identity \eqref{l2nu2}, we get
\begin{subequations}\label{Fll}
\begin{align}
{}_s {\cal F}_{\ell \,\ell- 1}^m  &\equiv {}_s\kappa_\ell^m \frac{\myzeta_{\ell}^m}{\myzeta_{\ell-1}^m}  \\
  &=- \ii[(\ell-1)/\ellc+m\ii/\ells] {}_s \bar \kappa_\ell^m \nonumber\\
{}_s {\cal F}_{\ell \,\ell+1}^m &\equiv {}_s\kappa_{\ell+1}^m  \frac{\myzeta_{\ell}^m}{\myzeta_{\ell+1}^m}  \\
  &=\ii    [(\ell+2)/\ellc-m\ii/\ells] {}_s \bar \kappa_{\ell+1}^m  \nonumber
\end{align}
\end{subequations}
where
\be
{}_s \bar \kappa_\ell^m  \equiv \sqrt{\frac{(\ell^2-s^2)(\ell^2-m^2)}{\ell^2}}\,.
\ee
With these identities, the coupling between multipoles with neighbor
values of $\ell$ take a simpler form, and the hierarchy has the structure
\bea
\partial_\eta \Theta_\ell^{m} &=& \sum_{\ell'=\ell\pm 1} {}_0 {\cal F}_{\ell
  \ell'}^m \Theta_{\ell'}^m +{\cal
  G}_\ell^m +{}^\Theta {\cal C}_\ell^m -\tau' \Theta_\ell^m\,,
\nonumber\\
\partial_\eta E_\ell^{m} &=& \sum_{\ell'=\ell\pm 1} {}_2{\cal F}_{\ell
  \ell'}^m E_{\ell'}^m - {\cal M}_\ell^m B_\ell^m \nonumber\\
&&\qquad +{}^E {\cal  C}_\ell^m -\tau' E_\ell^m\,, \label{PontzenHierarchy}\\
\partial_\eta B_\ell^{m} &=& \sum_{\ell'=\ell\pm 1} {}_2{\cal F}_{\ell
  \ell'}^m B_{\ell'}^m + {\cal M}_\ell^m  E_\ell^m -\tau' B_\ell^m\,,\nonumber
\eea
with
\be
 {\cal M}_\ell^m \equiv \frac{2 m \nu_m }{\ell(\ell+1)}
 =\frac{2}{\ell(\ell+1)}\left(\frac{m^2}{\ells}+\frac{\ii m}{\ellc}\right)\,.\nonumber
\ee
Up to variations in conventions, the hierarchy \eqref{PontzenHierarchy} is the same as the one obtained
in \cite{Pontzen2007}. Eventually, as we want to decompose the angular
dependence of the observed CMB directly on spin-weighted spherical
harmonic (without numerical prefactors) the CMB multipoles are related to the previous ones by
(7.35) of \cite{Pitrou:2019ifq}, that is
\be\label{ToCMB}
{}^{\rm CMB} \Theta_\ell^m(\eta_0) = \Theta_\ell^m(\eta_0) (-\ii)^\ell\sqrt{\frac{4\pi}{2\ell+1}}\,.
\ee
For a complete comparison with \cite{Pontzen2007} we also then need a parity inversion which brings factors of $(-1)^\ell$
(resp. $(-1)^{\ell+1}$) in the temperature and electric type multipoles (resp. the magnetic type multipoles) since those authors use multipoles defined with respect to the observed direction.

The difference between our method based on FLRW
  perturbations, and the method of \cite{Pontzen2007} based on Bianchi
  spaces directly, is manifest for the Bianchi $\VIIh$ and
  $\VIIo$. The invariant basis of these models are related to the
  invariant basis of the associated FLRW through \eqref{FiveToSevenh} or \eqref{FiveToSevenh2}. In the method of \cite{Pontzen2007}, one works fully in the basis $\gr{e}_i^{\VIIo}$ or $\gr{e}_i^{\VIIh}$. Hence the CMB sources are very simple at emission, as
  they are proportional to the matrices \eqref{Defqij}, but higher order multipoles
  are populated thanks the non trivial propagation of light in this
  basis. In our point of view, we work in the underlying FLRW basis
  $\gr{e}_i^{\I}$ (resp. $\gr{e}_i^{\V}$) when dealing with the case
  $\VIIo$ (resp. $\VIIh$). Hence the sources appear to have some large
  scale spiraling structure, but propagation is trivial as
  the direction of propagating photons is constant. To make it short,
  either the sources are simple but light propagation non-trivial as
  in \cite{Pontzen2007}, or sources are non-trivial but light
  propagation is simple, eventually leading to the same Boltzmann
  hierarchy. All these differences disappear for the Bianchi $\I$ and
  $\IX$ cases (see section \eqref{Simple19}) given their very simple structures.

\subsubsection{Special case of Bianchi $\I$ and $\IX$}\label{Simple19}

The case of Bianchi $\IX$ is much simpler than other models since we can see that there is no coupling to $\ell=1$ nor $\ell=3$. Of course we have $m=2$ so we need
at least $\ell\geq 2$ but we need also $\ell \leq 2$ because
$\nu=3/\ellc$. So we have a set of equations for the multipoles with
$\ell=2$ only which is
\bea\label{Boltzmann9}
\partial_\eta\Theta_2^{\pm 2} &=&{\cal G}_2^{\pm 2}+ {}^\Theta {\cal C}_2^{\pm 2}-\tau' \Theta_2^{\pm 2}\,,\\
\partial_\eta E_2^{\pm 2} &=& -2\ellc^{-1} B_2^{\pm 2}+ {}^E {\cal C}_2^{\pm 2}-\tau' E_2^{\pm 2}\,, \nonumber\\
\partial_\eta B_2^{\pm 2} &=& 2 \ellc^{-1}E_2^{\pm 2}-\tau' B_2^{\pm 2}\,.\nonumber
\eea
The radial functions needed for an integral solution are simply
\beas\label{Radial9}
{}_0 \bar \epsilon^{(2,2)}_2 &=& \frac{1}{5},\\
{}_2 \bar \epsilon^{(2,2)}_2 &=& \frac{1}{5}\cos(2 \chi/\ellc),\\
{}_2 \bar \beta^{(2,2)}_2 &=& \frac{1}{5} \sin(2 \chi/\ellc).
\eeas
Physically, polarization is rotated with respect to the invariant basis, and a quadrupole in $E$ (generated from scattering out of the
temperature quadrupole) converts to a quadrupole in $B$ and back~\cite{Pontzen2009}. 

The Bianchi $\I$ case is even simpler, and from the limit $\ellc \to
\infty$, $\ells \to \infty$ in \eqref{Fll}, we also check that there
are no couplings to $\ell \pm 1$, and magnetic type multipoles are not fed from free streaming of
electric type multipoles. Hence the system of equations is simply  
\bea\label{SimpleStructure}
\partial_\eta \Theta_2^{\pm m} &=&{\cal G}_2^{\pm m}+ {}^\Theta {\cal C}_2^{\pm m}-\tau' \Theta_2^{\pm m}\,,\\
\partial_\eta  E_2^{\pm m} &=& {}^E {\cal C}_2^{\pm m}-\tau' E_2^{\pm m}\,. \nonumber
\eea
Also the radial functions needed for the integral solutions
\eqref{LOS} are pure constants ($1/5$ for ${}_{s} \bar \epsilon_2^{(2m)}$) as seen from the limit $\ellc \to \infty$ in
\eqref{Radial9}, in agreement with the simple structure of the previous system which is trivially integrated on $\eta$.

\subsection{General cosmological observables}

All cosmological observables (weak lensing convergence or shear, lensing field, galaxy number counts,
redshift drifts, etc.) are of the form of an integral on the
background past light cone. However, when considering the effect of a Bianchi
perturbations, one must take into account the fact that it corresponds
to a pseudo plane-wave, and one must consider the effect of the
weights $\myzeta_\ell^m$. Quite similarly to the integral solution
\eqref{LOS} for CMB, one finds that the general solutions for
the multipoles of cosmological observables, when rephrased in terms of integrals on
sources multiplied by radial functions [Eq. (7.40) in \cite{Pitrou:2019ifq}],
need only be modified by
\be\label{RedefineAlpha}
{}_s \alpha_j^{(j'm)} \to \frac{\myzeta_{j}^m }{\myzeta_{j'}^m}\,{}_s \alpha_j^{(j'm)}\,.
\ee
Hence from the knowledge of the radial functions, which must be
computed from analytic continuation given that $\nu$ is complex for
some Bianchi perturbations, it is immediate to obtain theoretical
predictions for all observables using the same framework as the one
for stochastic linear perturbations.

In general, since we use the zenith direction for the reference axis, one
needs to allow for a general orientation of that Bianchi special
direction, as discussed in Section~\eqref{SecTransfo}. Rotations can be performed directly at the level of the
computed observables, that is rotating the angular multipoles of
observables. Similarly for models which are not invariant by a global
parity transformation ($\VIIo$, $\VIIh$ and $\IX$), we can perform the
parity transformation at the level of multipoles. For even type
multipoles (e.g. temperature or electric type polarization), this is a
factor $(-1)^\ell$ and for magnetic type ones a factor $(-1)^{\ell+1}$.

Finally, when computing multipoles, one can restrict the computation
to $m\geq 0$ since the negative $m$ are constrained by the fact that
the metric perturbation must be real [relations \eqref{qminusm} and \eqref{betasym}].
The CMB multipoles satisfy $X_\ell^{-m} = (-1)^m X_\ell^{m \star}$, with $X=\Theta,E,B$, and one encounters the same
relations for all cosmological observable multipoles.

\section{Conclusion}\label{Sec-Conclusions}

Bianchi models with isotropic limit, namely, models $\I$, $\VIIo$, $\VIIh$, $\V$ and $\IX$, are not alternative cosmologies. Rather, they are natural manifestations of linear and homogeneous cosmological perturbations in FLRW universes. The exact correspondence between nearly isotropic Bianchi 
and perturbed FLRW models allows for the computation of all angular multipoles of cosmological observables within the same FLRW framework. The modes $\nu_m$ required for 
the exact correspondence are summarized in Table \ref{TableMatching}, and our main results can be summarized as follows:
\begin{itemize}
  \item For models $\I,\IX$ the dynamics involves only $\ell=2$ multipoles, hence the
Boltzmann hierarchy is not really a hierarchy as it reduces to
\eqref{Boltzmann9} and \eqref{SimpleStructure}. Nonetheless one could use existing tools~\cite{CAMB,CLASS} for solving the
Einstein-Boltzmann set of equations, even though that would amount to
use a sledge hammer to kill a fly.
\item For the model $\VIIo$, the $\nu_m$ are real, and it corresponds to the effect of standing circularly polarized
  waves. Once again, existing tools are readily usable in that case.
\item However, since the correspondence of Bianchi models $\V$ and
  $\VIIh$ with perturbed FLRW is through super-curvature modes,  that is with complex modes $\nu_m$, one must rely on an
analytic continuation of the radial functions needed in the
expressions of the normal modes \eqref{DefGsjm}. In addition, the correspondence is not
with usual plane waves, but with pseudo plane-waves, which are
specified by the weights $\myzeta_\ell^m$ in the sum
\eqref{DefQfromsuml}. This requires to modify the usual Boltzmann
hierarchy of \cite{TAM2} with the rules \eqref{Modifiedkappa}. It also
modifies the integral solutions for any cosmological observable, but
this is simply equivalent to the redefinition \eqref{RedefineAlpha} for radial functions, as seen explicitly on the CMB case in Eqs.~\eqref{LOS}.
\end{itemize}

The power of the this approach is that one could compute the
non-stochastic part due to a Bianchi type perturbation with all the
sophistication of the usual linear perturbation theory around
FLRW spacetimes. For instance, that would allow to include consistently the anisotropic stress
of photons and neutrinos that should normally enter in the right hand
side of \eqref{betadynamic}. This method would avoid the
need to split numerical codes between a part dedicated to Bianchi
related effects, and another for stochastic perturbations, as done in \cite{Saadeh:2016bmp,Saadeh:2016sak}. Furthermore, using the integral solutions to
the CMB multipoles, that is the line of sight method of \cite{TAM1,TAM2}, would fasten the computations. Indeed,
exactly like it has allowed to solve the Boltzmann hierarchy of the stochastic component for a limited number of source multipoles, the
integral solution method would also allow to keep only a small number of multipoles when solving for
the sources instead of the full hierarchy (typically $\ell_{\rm
  max}=1000$ for the analysis of \cite{Saadeh:2016bmp}).
More importantly, this allows the possibility of computing
consistently the multipoles of various cosmological observables for
the same linear Bianchi perturbation, opening the possibility of joint
constraints on global anisotropy.

\acknowledgments

It is a pleasure to thank Oton Marcori and J-P. Uzan for useful
discussions during the initial stages of this work. TP thanks the
Institut d'Astrophysique de Paris for its hospitality, as well as
Brazilian Funding Agencies CNPq (311527/2018-3) and Fundação Araucária
(CP/PBA-2016) for their financial support.

\bibliography{BianchiDuality}

\begin{thebibliography}{41}
\expandafter\ifx\csname natexlab\endcsname\relax\def\natexlab#1{#1}\fi
\expandafter\ifx\csname bibnamefont\endcsname\relax
  \def\bibnamefont#1{#1}\fi
\expandafter\ifx\csname bibfnamefont\endcsname\relax
  \def\bibfnamefont#1{#1}\fi
\expandafter\ifx\csname citenamefont\endcsname\relax
  \def\citenamefont#1{#1}\fi
\expandafter\ifx\csname url\endcsname\relax
  \def\url#1{\texttt{#1}}\fi
\expandafter\ifx\csname urlprefix\endcsname\relax\def\urlprefix{URL }\fi
\providecommand{\bibinfo}[2]{#2}
\providecommand{\eprint}[2][]{\url{#2}}

\bibitem[{\citenamefont{Grishchuk et~al.}(1975)\citenamefont{Grishchuk,
  Doroshkevich, and Yudin}}]{Grishchuk:1975ec}
\bibinfo{author}{\bibfnamefont{L.~P.} \bibnamefont{Grishchuk}},
  \bibinfo{author}{\bibfnamefont{A.~G.} \bibnamefont{Doroshkevich}},
  \bibnamefont{and} \bibinfo{author}{\bibfnamefont{V.~M.} \bibnamefont{Yudin}},
  \bibinfo{journal}{Zh. Eksp. Teor. Fiz.} \textbf{\bibinfo{volume}{69}},
  \bibinfo{pages}{1857} (\bibinfo{year}{1975}).

\bibitem[{\citenamefont{King}(1991)}]{King:1991jd}
\bibinfo{author}{\bibfnamefont{D.~H.} \bibnamefont{King}},
  \bibinfo{journal}{Phys. Rev.} \textbf{\bibinfo{volume}{D44}},
  \bibinfo{pages}{2356} (\bibinfo{year}{1991}).

\bibitem[{\citenamefont{Pontzen and Challinor}(2011)}]{Pontzen2010}
\bibinfo{author}{\bibfnamefont{A.}~\bibnamefont{Pontzen}} \bibnamefont{and}
  \bibinfo{author}{\bibfnamefont{A.}~\bibnamefont{Challinor}},
  \bibinfo{journal}{Class. Quant. Grav.} \textbf{\bibinfo{volume}{28}},
  \bibinfo{pages}{185007} (\bibinfo{year}{2011}), \eprint{1009.3935}.

\bibitem[{\citenamefont{Pitrou and
  Pereira}(2019{\natexlab{a}})}]{DualityLetter}
\bibinfo{author}{\bibfnamefont{C.}~\bibnamefont{Pitrou}} \bibnamefont{and}
  \bibinfo{author}{\bibfnamefont{T.~S.} \bibnamefont{Pereira}}
  (\bibinfo{year}{2019}{\natexlab{a}}), \bibinfo{note}{in preparation}.

\bibitem[{\citenamefont{Adamek et~al.}(2016)\citenamefont{Adamek, Durrer, and
  Tansella}}]{Adamek:2015mna}
\bibinfo{author}{\bibfnamefont{J.}~\bibnamefont{Adamek}},
  \bibinfo{author}{\bibfnamefont{R.}~\bibnamefont{Durrer}}, \bibnamefont{and}
  \bibinfo{author}{\bibfnamefont{V.}~\bibnamefont{Tansella}},
  \bibinfo{journal}{JCAP} \textbf{\bibinfo{volume}{1601}}, \bibinfo{pages}{024}
  (\bibinfo{year}{2016}), \eprint{1510.01566}.

\bibitem[{\citenamefont{Marcori et~al.}(2018)\citenamefont{Marcori, Pitrou,
  Uzan, and Pereira}}]{Marcori:2018cwn}
\bibinfo{author}{\bibfnamefont{O.~H.} \bibnamefont{Marcori}},
  \bibinfo{author}{\bibfnamefont{C.}~\bibnamefont{Pitrou}},
  \bibinfo{author}{\bibfnamefont{J.-P.} \bibnamefont{Uzan}}, \bibnamefont{and}
  \bibinfo{author}{\bibfnamefont{T.~S.} \bibnamefont{Pereira}},
  \bibinfo{journal}{Phys. Rev.} \textbf{\bibinfo{volume}{D98}},
  \bibinfo{pages}{023517} (\bibinfo{year}{2018}), \eprint{1805.12121}.

\bibitem[{\citenamefont{Kodama and Sasaki}(1984)}]{Kodama:1985bj}
\bibinfo{author}{\bibfnamefont{H.}~\bibnamefont{Kodama}} \bibnamefont{and}
  \bibinfo{author}{\bibfnamefont{M.}~\bibnamefont{Sasaki}},
  \bibinfo{journal}{Prog. Theor. Phys. Suppl.} \textbf{\bibinfo{volume}{78}},
  \bibinfo{pages}{1} (\bibinfo{year}{1984}).

\bibitem[{\citenamefont{Mukhanov et~al.}(1992)\citenamefont{Mukhanov, Feldman,
  and Brandenberger}}]{Mukhanov:1990me}
\bibinfo{author}{\bibfnamefont{V.~F.} \bibnamefont{Mukhanov}},
  \bibinfo{author}{\bibfnamefont{H.~A.} \bibnamefont{Feldman}},
  \bibnamefont{and} \bibinfo{author}{\bibfnamefont{R.~H.}
  \bibnamefont{Brandenberger}}, \bibinfo{journal}{Phys. Rept.}
  \textbf{\bibinfo{volume}{215}}, \bibinfo{pages}{203} (\bibinfo{year}{1992}).

\bibitem[{\citenamefont{Barrow et~al.}(1985)\citenamefont{Barrow, Juszkiewicz,
  and Sonoda}}]{Barrow:1985tda}
\bibinfo{author}{\bibfnamefont{J.~D.} \bibnamefont{Barrow}},
  \bibinfo{author}{\bibfnamefont{R.}~\bibnamefont{Juszkiewicz}},
  \bibnamefont{and} \bibinfo{author}{\bibfnamefont{D.~H.}
  \bibnamefont{Sonoda}}, \bibinfo{journal}{Mon. Not. Roy. Astron. Soc.}
  \textbf{\bibinfo{volume}{213}}, \bibinfo{pages}{917} (\bibinfo{year}{1985}).

\bibitem[{\citenamefont{Jaffe et~al.}(2006)\citenamefont{Jaffe, Banday,
  Eriksen, Gorski, and Hansen}}]{Jaffe:2006sq}
\bibinfo{author}{\bibfnamefont{T.~R.} \bibnamefont{Jaffe}},
  \bibinfo{author}{\bibfnamefont{A.~J.} \bibnamefont{Banday}},
  \bibinfo{author}{\bibfnamefont{H.~K.} \bibnamefont{Eriksen}},
  \bibinfo{author}{\bibfnamefont{K.~M.} \bibnamefont{Gorski}},
  \bibnamefont{and} \bibinfo{author}{\bibfnamefont{F.~K.}
  \bibnamefont{Hansen}}, \bibinfo{journal}{Astron. Astrophys.}
  \textbf{\bibinfo{volume}{460}}, \bibinfo{pages}{393} (\bibinfo{year}{2006}),
  \eprint{astro-ph/0606046}.

\bibitem[{\citenamefont{Jaffe et~al.}(2005)\citenamefont{Jaffe, Banday,
  Eriksen, Gorski, and Hansen}}]{Jaffe:2005pw}
\bibinfo{author}{\bibfnamefont{T.~R.} \bibnamefont{Jaffe}},
  \bibinfo{author}{\bibfnamefont{A.~J.} \bibnamefont{Banday}},
  \bibinfo{author}{\bibfnamefont{H.~K.} \bibnamefont{Eriksen}},
  \bibinfo{author}{\bibfnamefont{K.~M.} \bibnamefont{Gorski}},
  \bibnamefont{and} \bibinfo{author}{\bibfnamefont{F.~K.}
  \bibnamefont{Hansen}}, \bibinfo{journal}{Astrophys. J.}
  \textbf{\bibinfo{volume}{629}}, \bibinfo{pages}{L1} (\bibinfo{year}{2005}),
  \eprint{astro-ph/0503213}.

\bibitem[{\citenamefont{McEwen et~al.}(2013)\citenamefont{McEwen, Josset,
  Feeney, Peiris, and Lasenby}}]{McEwen:2013cka}
\bibinfo{author}{\bibfnamefont{J.~D.} \bibnamefont{McEwen}},
  \bibinfo{author}{\bibfnamefont{T.}~\bibnamefont{Josset}},
  \bibinfo{author}{\bibfnamefont{S.~M.} \bibnamefont{Feeney}},
  \bibinfo{author}{\bibfnamefont{H.~V.} \bibnamefont{Peiris}},
  \bibnamefont{and} \bibinfo{author}{\bibfnamefont{A.~N.}
  \bibnamefont{Lasenby}}, \bibinfo{journal}{Mon. Not. Roy. Astron. Soc.}
  \textbf{\bibinfo{volume}{436}}, \bibinfo{pages}{3680} (\bibinfo{year}{2013}),
  \eprint{1303.3409}.

\bibitem[{\citenamefont{Pontzen and Challinor}(2007)}]{Pontzen2007}
\bibinfo{author}{\bibfnamefont{A.}~\bibnamefont{Pontzen}} \bibnamefont{and}
  \bibinfo{author}{\bibfnamefont{A.}~\bibnamefont{Challinor}},
  \bibinfo{journal}{Mon. Not. Roy. Astron. Soc.}
  \textbf{\bibinfo{volume}{380}}, \bibinfo{pages}{1387} (\bibinfo{year}{2007}),
  \eprint{0706.2075}.

\bibitem[{\citenamefont{Pontzen}(2009)}]{Pontzen2009}
\bibinfo{author}{\bibfnamefont{A.}~\bibnamefont{Pontzen}},
  \bibinfo{journal}{Phys. Rev.} \textbf{\bibinfo{volume}{D79}},
  \bibinfo{pages}{103518} (\bibinfo{year}{2009}), \eprint{0901.2122}.

\bibitem[{\citenamefont{Sung and Coles}(2011)}]{Sung:2010ek}
\bibinfo{author}{\bibfnamefont{R.}~\bibnamefont{Sung}} \bibnamefont{and}
  \bibinfo{author}{\bibfnamefont{P.}~\bibnamefont{Coles}},
  \bibinfo{journal}{JCAP} \textbf{\bibinfo{volume}{1106}}, \bibinfo{pages}{036}
  (\bibinfo{year}{2011}), \eprint{1004.0957}.

\bibitem[{\citenamefont{Saadeh et~al.}(2016{\natexlab{a}})\citenamefont{Saadeh,
  Feeney, Pontzen, Peiris, and McEwen}}]{Saadeh:2016bmp}
\bibinfo{author}{\bibfnamefont{D.}~\bibnamefont{Saadeh}},
  \bibinfo{author}{\bibfnamefont{S.~M.} \bibnamefont{Feeney}},
  \bibinfo{author}{\bibfnamefont{A.}~\bibnamefont{Pontzen}},
  \bibinfo{author}{\bibfnamefont{H.~V.} \bibnamefont{Peiris}},
  \bibnamefont{and} \bibinfo{author}{\bibfnamefont{J.~D.}
  \bibnamefont{McEwen}}, \bibinfo{journal}{Mon. Not. Roy. Astron. Soc.}
  \textbf{\bibinfo{volume}{462}}, \bibinfo{pages}{1802}
  (\bibinfo{year}{2016}{\natexlab{a}}), \eprint{1604.01024}.

\bibitem[{\citenamefont{Saadeh et~al.}(2016{\natexlab{b}})\citenamefont{Saadeh,
  Feeney, Pontzen, Peiris, and McEwen}}]{Saadeh:2016sak}
\bibinfo{author}{\bibfnamefont{D.}~\bibnamefont{Saadeh}},
  \bibinfo{author}{\bibfnamefont{S.~M.} \bibnamefont{Feeney}},
  \bibinfo{author}{\bibfnamefont{A.}~\bibnamefont{Pontzen}},
  \bibinfo{author}{\bibfnamefont{H.~V.} \bibnamefont{Peiris}},
  \bibnamefont{and} \bibinfo{author}{\bibfnamefont{J.~D.}
  \bibnamefont{McEwen}}, \bibinfo{journal}{Phys. Rev. Lett.}
  \textbf{\bibinfo{volume}{117}}, \bibinfo{pages}{131302}
  (\bibinfo{year}{2016}{\natexlab{b}}), \eprint{1605.07178}.

\bibitem[{\citenamefont{Pitrou and
  Pereira}(2019{\natexlab{b}})}]{Pitrou:2019ifq}
\bibinfo{author}{\bibfnamefont{C.}~\bibnamefont{Pitrou}} \bibnamefont{and}
  \bibinfo{author}{\bibfnamefont{T.~S.} \bibnamefont{Pereira}},
  \bibinfo{journal}{Phys. Rev. D} \textbf{\bibinfo{volume}{100}},
  \bibinfo{pages}{123535} (\bibinfo{year}{2019}{\natexlab{b}}),
  \eprint{1909.13687}.

\bibitem[{\citenamefont{Ellis et~al.}(2012)\citenamefont{Ellis, Maartens, and
  MacCallum}}]{ellis2012relativistic}
\bibinfo{author}{\bibfnamefont{G.~F.} \bibnamefont{Ellis}},
  \bibinfo{author}{\bibfnamefont{R.}~\bibnamefont{Maartens}}, \bibnamefont{and}
  \bibinfo{author}{\bibfnamefont{M.~A.} \bibnamefont{MacCallum}},
  \emph{\bibinfo{title}{Relativistic cosmology}} (\bibinfo{publisher}{Cambridge
  University Press}, \bibinfo{year}{2012}).

\bibitem[{\citenamefont{Plebanski and
  Krasinski}(2006)}]{plebanski2006introduction}
\bibinfo{author}{\bibfnamefont{J.}~\bibnamefont{Plebanski}} \bibnamefont{and}
  \bibinfo{author}{\bibfnamefont{A.}~\bibnamefont{Krasinski}},
  \emph{\bibinfo{title}{An introduction to general relativity and cosmology}}
  (\bibinfo{publisher}{Cambridge University Press}, \bibinfo{year}{2006}).

\bibitem[{\citenamefont{{\O}yvind and Hervik}(2007)}]{oyvind2007einstein}
\bibinfo{author}{\bibfnamefont{G.}~\bibnamefont{{\O}yvind}} \bibnamefont{and}
  \bibinfo{author}{\bibfnamefont{S.}~\bibnamefont{Hervik}},
  \emph{\bibinfo{title}{Einstein's General Theory of Relativity: With Modern
  Applications in Cosmology}} (\bibinfo{publisher}{Springer},
  \bibinfo{year}{2007}).

\bibitem[{\citenamefont{Ellis and MacCallum}(1969)}]{Ellis1969}
\bibinfo{author}{\bibfnamefont{G.~F.~R.} \bibnamefont{Ellis}} \bibnamefont{and}
  \bibinfo{author}{\bibfnamefont{M.~A.~H.} \bibnamefont{MacCallum}},
  \bibinfo{journal}{Communications in Mathematical Physics}
  \textbf{\bibinfo{volume}{12}}, \bibinfo{pages}{108} (\bibinfo{year}{1969}),
  ISSN \bibinfo{issn}{1432-0916},
  \urlprefix\url{https://doi.org/10.1007/BF01645908}.

\bibitem[{\citenamefont{{Collins} and
  {Hawking}}(1973{\natexlab{a}})}]{Collins1973a}
\bibinfo{author}{\bibfnamefont{C.~B.} \bibnamefont{{Collins}}}
  \bibnamefont{and} \bibinfo{author}{\bibfnamefont{S.~W.}
  \bibnamefont{{Hawking}}}, \bibinfo{journal}{\apj}
  \textbf{\bibinfo{volume}{180}}, \bibinfo{pages}{317}
  (\bibinfo{year}{1973}{\natexlab{a}}).

\bibitem[{\citenamefont{Barrow and Hervik}(2003)}]{Barrow:2003fc}
\bibinfo{author}{\bibfnamefont{J.~D.} \bibnamefont{Barrow}} \bibnamefont{and}
  \bibinfo{author}{\bibfnamefont{S.}~\bibnamefont{Hervik}},
  \bibinfo{journal}{Class. Quant. Grav.} \textbf{\bibinfo{volume}{20}},
  \bibinfo{pages}{2841} (\bibinfo{year}{2003}), \eprint{gr-qc/0304050}.

\bibitem[{\citenamefont{Ellis and van Elst}(1999)}]{Ellis1998}
\bibinfo{author}{\bibfnamefont{G.~F.~R.} \bibnamefont{Ellis}} \bibnamefont{and}
  \bibinfo{author}{\bibfnamefont{H.}~\bibnamefont{van Elst}},
  \bibinfo{journal}{NATO Sci. Ser. C} \textbf{\bibinfo{volume}{541}},
  \bibinfo{pages}{1} (\bibinfo{year}{1999}), \eprint{gr-qc/9812046}.

\bibitem[{\citenamefont{Tsagas et~al.}(2008)\citenamefont{Tsagas, Challinor,
  and Maartens}}]{Tsagas:2007yx}
\bibinfo{author}{\bibfnamefont{C.~G.} \bibnamefont{Tsagas}},
  \bibinfo{author}{\bibfnamefont{A.}~\bibnamefont{Challinor}},
  \bibnamefont{and} \bibinfo{author}{\bibfnamefont{R.}~\bibnamefont{Maartens}},
  \bibinfo{journal}{Phys. Rept.} \textbf{\bibinfo{volume}{465}},
  \bibinfo{pages}{61} (\bibinfo{year}{2008}), \eprint{0705.4397}.

\bibitem[{\citenamefont{Gourgoulhon}(2007)}]{Gourgoulhon:2007ue}
\bibinfo{author}{\bibfnamefont{E.}~\bibnamefont{Gourgoulhon}}
  (\bibinfo{year}{2007}), \eprint{gr-qc/0703035}.

\bibitem[{\citenamefont{Pitrou et~al.}(2013)\citenamefont{Pitrou, Roy, and
  Umeh}}]{xPand}
\bibinfo{author}{\bibfnamefont{C.}~\bibnamefont{Pitrou}},
  \bibinfo{author}{\bibfnamefont{X.}~\bibnamefont{Roy}}, \bibnamefont{and}
  \bibinfo{author}{\bibfnamefont{O.}~\bibnamefont{Umeh}},
  \bibinfo{journal}{Class. Quant. Grav.} \textbf{\bibinfo{volume}{30}},
  \bibinfo{pages}{165002} (\bibinfo{year}{2013}), \eprint{1302.6174}.

\bibitem[{\citenamefont{{Collins} and
  {Hawking}}(1973{\natexlab{b}})}]{Collins1973b}
\bibinfo{author}{\bibfnamefont{C.~B.} \bibnamefont{{Collins}}}
  \bibnamefont{and} \bibinfo{author}{\bibfnamefont{S.~W.}
  \bibnamefont{{Hawking}}}, \bibinfo{journal}{MNRAS}
  \textbf{\bibinfo{volume}{162}}, \bibinfo{pages}{307}
  (\bibinfo{year}{1973}{\natexlab{b}}).

\bibitem[{\citenamefont{Ma and Bertschinger}(1995)}]{Ma1995}
\bibinfo{author}{\bibfnamefont{C.-P.} \bibnamefont{Ma}} \bibnamefont{and}
  \bibinfo{author}{\bibfnamefont{E.}~\bibnamefont{Bertschinger}},
  \bibinfo{journal}{Astrophys. J.} \textbf{\bibinfo{volume}{455}},
  \bibinfo{pages}{7} (\bibinfo{year}{1995}), \eprint{astro-ph/9506072}.

\bibitem[{\citenamefont{Peter and Uzan}(2005)}]{pubook}
\bibinfo{author}{\bibfnamefont{P.}~\bibnamefont{Peter}} \bibnamefont{and}
  \bibinfo{author}{\bibfnamefont{J.-P.} \bibnamefont{Uzan}},
  \emph{\bibinfo{title}{{Primordial Cosmology}}} (\bibinfo{publisher}{Oxford
  University Press}, \bibinfo{year}{2005}).

\bibitem[{\citenamefont{Malik and Wands}(2009)}]{Malik:2008im}
\bibinfo{author}{\bibfnamefont{K.~A.} \bibnamefont{Malik}} \bibnamefont{and}
  \bibinfo{author}{\bibfnamefont{D.}~\bibnamefont{Wands}},
  \bibinfo{journal}{Phys. Rept.} \textbf{\bibinfo{volume}{475}},
  \bibinfo{pages}{1} (\bibinfo{year}{2009}), \eprint{0809.4944}.

\bibitem[{\citenamefont{Hu and White}(1997)}]{TAM1}
\bibinfo{author}{\bibfnamefont{W.}~\bibnamefont{Hu}} \bibnamefont{and}
  \bibinfo{author}{\bibfnamefont{M.~J.} \bibnamefont{White}},
  \bibinfo{journal}{Phys. Rev.} \textbf{\bibinfo{volume}{D56}},
  \bibinfo{pages}{596} (\bibinfo{year}{1997}), \eprint{astro-ph/9702170}.

\bibitem[{\citenamefont{Hu et~al.}(1998)\citenamefont{Hu, Seljak, White, and
  Zaldarriaga}}]{TAM2}
\bibinfo{author}{\bibfnamefont{W.}~\bibnamefont{Hu}},
  \bibinfo{author}{\bibfnamefont{U.}~\bibnamefont{Seljak}},
  \bibinfo{author}{\bibfnamefont{M.~J.} \bibnamefont{White}}, \bibnamefont{and}
  \bibinfo{author}{\bibfnamefont{M.}~\bibnamefont{Zaldarriaga}},
  \bibinfo{journal}{Phys. Rev.} \textbf{\bibinfo{volume}{D57}},
  \bibinfo{pages}{3290} (\bibinfo{year}{1998}), \eprint{astro-ph/9709066}.

\bibitem[{\citenamefont{Lyth and Woszczyna}(1995)}]{Lyth1995}
\bibinfo{author}{\bibfnamefont{D.~H.} \bibnamefont{Lyth}} \bibnamefont{and}
  \bibinfo{author}{\bibfnamefont{A.}~\bibnamefont{Woszczyna}},
  \bibinfo{journal}{Phys. Rev.} \textbf{\bibinfo{volume}{D52}},
  \bibinfo{pages}{3338} (\bibinfo{year}{1995}), \eprint{astro-ph/9501044}.

\bibitem[{\citenamefont{Barrow and Levin}(1997)}]{Barrow:1997vu}
\bibinfo{author}{\bibfnamefont{J.~D.} \bibnamefont{Barrow}} \bibnamefont{and}
  \bibinfo{author}{\bibfnamefont{J.~J.} \bibnamefont{Levin}},
  \bibinfo{journal}{Phys. Lett.} \textbf{\bibinfo{volume}{A233}},
  \bibinfo{pages}{169} (\bibinfo{year}{1997}), \eprint{astro-ph/9704041}.

\bibitem[{\citenamefont{Stewart}(1990)}]{Stewart:1990fm}
\bibinfo{author}{\bibfnamefont{J.~M.} \bibnamefont{Stewart}},
  \bibinfo{journal}{Class. Quant. Grav.} \textbf{\bibinfo{volume}{7}},
  \bibinfo{pages}{1169} (\bibinfo{year}{1990}).

\bibitem[{\citenamefont{Pereira et~al.}(2016)\citenamefont{Pereira, Pitrou, and
  Uzan}}]{Pereira:2015jya}
\bibinfo{author}{\bibfnamefont{T.~S.} \bibnamefont{Pereira}},
  \bibinfo{author}{\bibfnamefont{C.}~\bibnamefont{Pitrou}}, \bibnamefont{and}
  \bibinfo{author}{\bibfnamefont{J.-P.} \bibnamefont{Uzan}},
  \bibinfo{journal}{Astron. Astrophys.} \textbf{\bibinfo{volume}{585}},
  \bibinfo{pages}{L3} (\bibinfo{year}{2016}), \eprint{1503.01127}.

\bibitem[{\citenamefont{Lewis and Challinor}(1999)}]{CAMB}
\bibinfo{author}{\bibfnamefont{A.}~\bibnamefont{Lewis}} \bibnamefont{and}
  \bibinfo{author}{\bibfnamefont{A.}~\bibnamefont{Challinor}},
  \emph{\bibinfo{title}{\mbox{CAMB}}},
  \bibinfo{howpublished}{\url{http://camb.info}} (\bibinfo{year}{1999}).

\bibitem[{\citenamefont{Lesgourgues}(2011)}]{CLASS}
\bibinfo{author}{\bibfnamefont{J.}~\bibnamefont{Lesgourgues}},
  \emph{\bibinfo{title}{\mbox{CLASS}}},
  \bibinfo{howpublished}{\url{http://class-code.net/}} (\bibinfo{year}{2011}).

\bibitem[{\citenamefont{Challinor}(2000)}]{Challinor:1999xz}
\bibinfo{author}{\bibfnamefont{A.}~\bibnamefont{Challinor}},
  \bibinfo{journal}{Class. Quant. Grav.} \textbf{\bibinfo{volume}{17}},
  \bibinfo{pages}{871} (\bibinfo{year}{2000}), \eprint{astro-ph/9906474}.

\end{thebibliography}

\appendix

\section{$1+3$ splitting of Einstein equations}\label{1p3review}

The Gauss-Codazzi relation is \cite{Gourgoulhon:2007ue}
\begin{eqnarray}\label{EqGaussCodazzi}
	 R_{\mu\nu\lambda\sigma} &= 
		& \, {\Rspatial}_{\mu\nu\lambda\sigma} + 2 K_{\mu [\lambda} K_{\sigma ] \nu}
		\\
&-& 4 \left( D_{[ \lambda} K_{\sigma] [\mu} \right) e_{\nu ]} 
		+ 4 \, e_{[ \mu} \, K_{\nu ]}^{\phantom{b} \rho} \,
                K_{\rho [ \lambda} \, e_{\sigma ]}\nonumber\\
&+& 4 e_{[\mu} \dot{{K}}_{\nu] [ \lambda} e_{\sigma ]} 	- 4 \left(  D_{[ \mu} K_{\nu ] [ \lambda} \right) e_{\sigma ]}\, , \nonumber
\end{eqnarray}
where a dot derivative stands for $e^\mu \nabla_\mu$.  

For a homogeneous projected tensor, one has \cite{xPand}
\begin{equation}\label{DTijkl}
D_k T_{i_1\dots i_p} = -\sum_{j=1}^p {\Gamma^{l}}_{k\, {i_j}} T_{i_1 \dots i_{j-1} \,l  \,i_{j+1} \dots i_p}\,.
\end{equation}
Hence, using the definition of the Riemann tensor from the commutation
of two such derivatives, we infer after using \eqref{GCIJK}, the
Riemann tensor associated with the spatial metric
\begin{eqnarray}\label{EqRiemannh} 
	{}^{(3)}R_{ij}^{\phantom{ij} kl} &=& - \frac{1}{2}{C^p}_{ij} {C_p}^{kl} + \frac{1}{2}C_{p\phantom{d}i}^{\phantom{e}l} {C^{pk}}_j 
		+ C_{p\phantom{d}j}^{\phantom{p}l} {C_i}^{kp}
		\nonumber\\
&&+ C_{p\phantom{d}j}^{\phantom{e}l} C^{k\phantom{b}p}_{\phantom{c}i} 
		+ {C}_{ijp} {C}^{pkl} 
		+ \frac{1}{2}C_{i\phantom{d}p}^{\phantom{a}l}
                {C_j}^{kp}\nonumber\\
& &+ \frac{1}{2}{C^l}_{ip} C^{k\phantom{b}p}_{\phantom{c}j} 
		+ {C^k}_{jp} {C_i}^{lp} \, .
\end{eqnarray}
The three-Ricci tensor and three-Ricci scalar can then be deduced, and we obtain:
\begin{align}
	\Rspatial_{ij} = 
		& - \frac{1}{2} C_{kil} C^{k\phantom{j}l}_{\phantom{k}j}
		- \frac{1}{2} C_{kil} C^{l\phantom{j}k}_{\phantom{l}j}
                  \nonumber\\
&
		+ \frac{1}{4} {C_i}^{kl} C_{jkl} 
		+ {C_{(ij)}}^p {C^k}_{pk} \, , \label{eq:RicciToConstants} \\
	\Rspatial = 
		& - \frac{1}{4} C_{ijk}C^{ijk} 
		- \frac{1}{2} C_{ijk} C^{jik} 
		+ {C^{kj}}_k {C^p}_{pj}\, . \label{eq:RicciScalToConstants}
\end{align}
Whenever the placement of indices on the constants of structure is not
${C^i}_{jk}$, it implies that indices are either lowered by $h_{ij}$ or raised by $h^{ij}$.  In particular using the general decomposition \eqref{DefNA} the Ricci and Ricci scalar are given
\bea
\Rspatial_{ij} &=& h_{ij} \left[-2A_k A^k -N_{kl}N^{kl}
  +\tfrac{1}{2}(N_k^k)^2\right]\nonumber\\
&&+2 {N_{i}}^k N_{kj} - N_{ij} N_k^k\nonumber\\
&&-2 \epsilon_{kl(j} {N_{i)}}^l A^k \,,
\eea
\be\label{EqR3}
\Rspatial = -6 A_i A^i -N_{ij} N^{ij} +\tfrac{1}{2}(N_i^i)^2\,.
\ee
Note that the traceless part of the Ricci scalar is 
\bea
\Rspatial_{\langle ij \rangle} &=& - N_{\langle ij \rangle} N_k^k +
2{N_{\langle i}}^k N_{j\rangle k}\nonumber\\
&&-2 \epsilon_{kl(j} {N_{i)}}^l A^k\,,
\eea
and it vanishes whenever $N^{ij} = 0$.

From \eqref{DTijkl} we find that for spatial (projected) symmetric
trace-free tensors $T^{\mu_1\dots \mu_n}$ 
\beas\label{UsefulDVDT1}
D^k T_{k i_1 \dots i_n} &=& - (n+2) A^k T_{k i_1 \dots i_n}\nonumber\\
&&-n \epsilon^j_{\,\,l \langle i_1} T_{i_2 \dots i_n \rangle kj}
N^{lk}\,,\\
{\rm curl}\, T_{i_1 \dots i_n} &=&-A^j \epsilon_{jk\langle
  i_1}{T_{i_2\dots i_n\rangle}}^k\\
&&-\frac{(n-1)}{2} N^k_k T_{i_1\dots i_n}\nonumber\\
&&+(2n-1) N^k_{\langle i_1}T_{i_2 \dots i_n \rangle k}\,, \slabel{CurlSFT}\nonumber\\
D_{\langle j} T_{i_1 \dots i_n \rangle} &=&n A_{\langle j} T_{i_1
  \dots i_n \rangle} \nonumber\\
&&-n \epsilon_{lk\langle i_1} {T^l}_{i_2 \dots i_n} N^k_{j \rangle}\,,\slabel{DSTFT}
\eeas
where we have introduced the curl in curved space
\be\label{DefCurl}
{\rm curl}\, T_{ij\dots k} = \epsilon_{rs(i}D^r {T_{j\dots k)}}^s\,.
\ee
In particular for a homogeneous projector vector $V^\mu$, and a symmetric trace-free homogeneous projected tensor $T^{\mu\nu}$ 
\bea\label{UsefulDVDT}
D_j V^j &=& - 2 A_i V^i\,,\\
D_j T^{j}_{\,i} &=&-3 A_j T^{j}_{\,i} -\epsilon_{i jk}
T^{j}_{\,p} N^{p k}\,.
\eea
This last relation can be used for spatial derivatives on extrinsic
curvature (on separating its trace part) in
\eqref{EqGaussCodazzi}. Furthermore, the Gauss-Codazzi relation is
used in practice by also converting the dot derivative to Lie derivative on the extrinsic curvature. In general, the dot derivative of a homogeneous tensor is transformed to a Lie derivative using
\begin{equation}
	{\cal L}_{\gr{e}}  T_{i_1 \dots i_p} 
		= \dot{T}_{i_1 \dots i_p} 
		+ \sum_{i=1}^{p}  K^{j}_{\phantom{b} i_i} \, T_{i_1 \dots i_{i-1} j i_{i+1} \dots i_p} \, .
\end{equation}
We need quite often the relations
\beas\label{DotToLie}
{\cal L}_{\gr{e}} h_{ij} &=& \frac{2}{3} \theta h_{ij} + 2 \sigma_{ij}\,,\\
 {\cal L}_{\gr{e}} h^{ij} &=& -\frac{2}{3} \theta h^{ij} -2 \sigma^{ij}\,,\\
{\cal L}_{\gr{e}} \sigma_{ij} &=& \dot \sigma_{ij} + \frac{2}{3}\theta
\sigma_{ij} + 2 \sigma_{ikc}{\sigma^{k}}_j\,,\\
{\cal L}_{\gr{e}} \sigma^{i}_j &=& \dot \sigma^i_{j}\,,\\
 {\cal L}_{\gr{e}} \sigma^{ij} &=& \dot \sigma^{ij} - \frac{2}{3}\theta
\sigma^{ij} - 2 \sigma^{ik}{\sigma_{k}}^j\,.
\eeas

A contraction of the Gauss-Codazzi relation leads to the Raychaudhuri equation
\begin{equation}\label{Raychaudhuri}
 \dot\theta=-\frac{1}{3}\theta^2-\sigma^2 - R_{\mu\nu} e^\mu e^\nu, \,\, {\rm
   with}\,\, \sigma^2 \equiv \sigma_{ij}\sigma^{ij}
\end{equation}
The  general Friedmann equation (constraint) is another contraction of the Gauss-Codazzi relation and is
\begin{equation}\label{GaussCodazziR}
\Rspatial = -\frac{2}{3}\theta^2+\sigma^2+2 G_{\mu\nu} e^\mu e^\nu.
 \end{equation}

The tilt constraint is found for yet another contraction leading to
\begin{equation}
h_i^k e^\mu G_{\mu k} =D_j K^j_i=-3 A^k \sigma_{kj} -\epsilon_{jkp} \sigma^{kq} N_q^p\,.
\end{equation}
And finally for the shear evolution we get
\begin{eqnarray}
\dot \sigma_{ij} + \theta \sigma_{ij} &=& N_{\langle ij \rangle} N^k_k
-2 N_{\langle i}^k N_{j\rangle k} +2 A^k \epsilon_{kp\langle i} N_{j
  \rangle}^p\nonumber\\
&&+ G_{\langle i j\rangle}\,.
\end{eqnarray}
It can be easily recast with a Lie derivative using Eqs.~\eqref{DotToLie}.

\section{Open case coordinates and basis}\label{AppOpenCoordinates}

Here we give expressions for the KVFs and invariant basis of the
Bianchi~$\V$ and $\VIIh$ model using the 
$(x,y,z)$ coordinates of Ref.~\cite{Barrow:1997vu}. These are related to the spherical hyperbolic
coordinates in~\eqref{openmetricspherical} by
\begin{align}
x & =  \ellc \exp(+z/\ellc)\sinh(\chi/\ellc) \sin \theta \cos \phi\,,\nonumber\\
y & =  \ellc \exp(+z/\ellc)\sinh(\chi/\ellc) \sin \theta \sin \phi\,,\label{barrow-y}\\
z & =  -\ellc \ln[\cosh(\chi/\ellc)\! -\! \sinh(\chi/\ellc) \cos \theta]\nonumber\,.
\end{align}
In terms of these variables, the metric of the open space becomes
\be\label{openmetricxyz}
\dd s^2 = \dd z^2 + \exp(-2 z/\ellc)(\dd x^2 + \dd y^2)\,.
\ee
Because these variables are more adapted to the symmetries of the Bianchi~$\V$ models, the KVFs and
invariant basis simplify considerably. The KVFs are
\bea
\gr{\xi}^{\V}_x &=&  \partial_x\,,\nonumber\\
\gr{\xi}^{\V}_y &=&  \partial_y\,,\\
\gr{\xi}^{\V}_z &=&  (x/\ellc) \partial_x +(y/\ellc)  \partial_y + \partial_z\,.\nonumber
\eea
These solve~\eqref{ExplicitCommuteXi} with $N^i=0$, as one can check. The corresponding invariant basis is given by
\bea\label{InvbasisV}
\gr{e}_x^{\V}&=& \exp(z/\ellc) \partial_x\,,\nonumber\\
\gr{e}_y^{\V}&=& \exp(z/\ellc) \partial_y\,,\\
\gr{e}_z^{\V}&=& \partial_z\,.\nonumber
\eea
As for the $\VIIh$ case, the KVFs are
	\bea
	\gr{\xi}^{\VIIh}_1 &=&  \partial_x\nonumber\\
	\gr{\xi}^{\VIIh}_2 &=&  \partial_y\\
	\gr{\xi}^{\VIIh}_3 &=&  \left(\frac{x}{\ellc} +
	\frac{y}{\ells}\right) \partial_x + \left(\frac{y}{\ellc}- \frac{x}{\ells}\right)  \partial_y + \partial_z\,,\nonumber
      \eea
whereas the invariant basis is trivially found from~\eqref{InvbasisV} and \eqref{FiveToSevenh}.

\section{Invariant co-basis}\label{AppCoBasis}

In any case, denoting $\mu,\nu \dots$ components in basis or co-basis
associated with coordinates ($x,y,z$ or $\chi,\theta,\phi$), the
invariant co-basis of a given Bianchi type is related to the invariant basis through 
\be
e^{i}_{\,\,\mu} = g^{ij} g_{\mu\nu} e_j^{\,\,\nu}\,.
\ee
Here $g^{ij}$ are the components of the inverse metric in the
invariant co-basis, and $g_{\mu\nu}$ the components of the metric in
the coordinate basis. For Bianchi types $\I,\V,\IX$, by construction
$g^{ij}=\delta^{ij}$. Hence finding the co-basis from the basis is
straightforward. For instance from the invariant basis in spherical
coordinates in the Bianchi $\V$ case \ref{InvbasisV}, one infers
trivially the invariant co-basis. For the Bianchi $\VIIh$ case, since
the invariant basis is related to the one of $\V$ by \eqref{FiveToSevenh}, finding the invariant co-basis of $\VIIh$ from the
one of $\V$ is also straightforward from \eqref{FiveToSevenh2}. 

\section{Matching equations (finding the $\nu_m$)}\label{MatchEquations}

In this section we gather all the details of the identification
FLRW-Bianchi at the level of equations, allowing to determine which modes $\nu_m$ are
needed to realize the matching~\eqref{qQplus}.

\subsection{Tensors}

The matching between tensor modes is the easiest to make. For that we just need to look for tensors such that the definition
\be\label{EijToqij1}
E^{(\pm2)}_{ij} \equiv \beta_{(\pm2)} q^{(\pm2)}_{ij}\,
\ee
holds. The two main equations in this case are Eqs.~\eqref{TensorEqFL} and~\eqref{trace-free-eq}. Their direct comparison leads to
\be
{\cal S}^{(\pm2)} = -\nu_{\pm2}^2 + K\,,
\ee
where we have used~\eqref{laplace-beltrami} with $j=|m|=2$. Using the
results of Table~\ref{svtTable}, it is now straightforward to find the
mode $\nu_{\pm 2}$ which connects these two equations. For flat ($K=0$) models we find
\be
\nu_{\pm 2}=\begin{cases}
	0\,, & (\I)\,,\\
	\pm\frac{2}{\ells}\,, & (\VIIo)\,.
\end{cases}
\ee
For open models, $K=-\ellc^{-2}$ and we find
\be
\nu_{\pm 2}=\begin{cases}
	\frac{\ii}{\ellc} \,, & (\V)\,,\\
	\pm\frac{2}{\ells}+\frac{\ii}{\ellc}\,, & (\VIIh)\,.
\end{cases}
\ee
Note that we can obtain models $\I$ and $\V$ as limits of models $\VIIo$ and $\VIIh$, respectively, by taking $\ells$ to infinity while keeping $\ellc$ fixed (see Eq.~\eqref{Defsqrth}).

For the closed model ($K=\ellc^{-2}$) we have
\be
\nu_{\pm 2} = \pm \frac{3}{\ellc} \qquad (\IX)\,.
\ee
or $|k|=\sqrt{6}\ellc^{-1}$. One can show that this corresponds to a tensor wave whose length equals one half the curvature radius of a closed  universe~\cite{King:1991jd}. Finally, we note that in all cases the momentum constraint ${\cal P}^{(m)}_i=0$ is obvious, since tensor perturbations do not induce momentum.

\subsection{Vectors}

In principle the matching of the vector modes follows similarly, that is, we introduce modes $E^{(\pm1)}_i$ satisfying
\be\label{EijToqij2}
\hatD_{(i} E^{(\pm1)}_{j)} \equiv \beta_{(\pm1)} q^{(\pm1)}_{ij}\,.
\ee
and then use it to compare Eqs.~\eqref{2nd-v-constraint}
and~\eqref{trace-free-eq}, from which we deduce the $\nu_{\pm 1}$ with the help of~\eqref{laplace-beltrami}. But since the former does not contain a Laplacian term, and the latter has ${\cal S}^{ij}=0$ for flat and open models (the closed case is discussed below), this comparison will lead us nowhere. 

We can instead find $\nu_{\pm 1}$ by comparing the constraint equation~\eqref{2nd-v-constraint} with the tilt~\eqref{DefTilt}, provided we have an explicit solution to Eq.~\eqref{EijToqij2}. Such solution can be constructed by writing $E^{(\pm1)}_i$ as a linear combination of the invariant basis and using Eq.~\eqref{DTijkl} to fix the coefficients. Since ${\cal D}_{(i}e_{j)}=0$ in model $\I$ (see~\S\ref{KVF-and-invbasis}), non-trivial solutions 
can only be constructed in models $\VIIo$, $\VIIh$ and $\V$. As it turns out, a solution to these models can be written jointly as~\cite{Pontzen2010}
\be\label{Epm1-solution}
E^{(\pm1)}_i = \pm\ells\beta_{(\pm1)}\left(\frac{\ells/\ellc\mp\ii}{1+(\ells/\ellc)^2}\right)(e^{(\mp)}_\VIIh)_i\,.
\ee
One can easily check that model $\VIIo$ is recovered when $\ellc \to \infty$ (and thus $h \to 0$), and model 
$\V$ in the limit $\ells\rightarrow\infty$ (and thus $h \to \infty$).

If we now use~\eqref{Pi-svt} to decompose the right-hand side of~\eqref{2nd-v-constraint} and use~\eqref{laplace-beltrami} with $j=|m|=1$, we find
\be\label{tilt-matching}
\beta'_{(\pm1)}{\cal P}^{(\pm1)}_i = \left(\frac{(\nu_{\pm 1})^2}{2} - 2K \right) (E^{(\pm 1)}_i )'\,.
\ee
This, together with~\eqref{Epm1-solution} and the results of
Table~\ref{svtTable}, allows us to find the $\nu_{\pm 1}$ for each model. For flat models this gives
\be
\nu_{\pm 1}=\begin{cases}\label{nuvectorflat}
	0 \,, & (\I)\,,\\
	\pm \frac{1}{\ells}\,, & (\VIIo)\,.
\end{cases}
\ee
For open models, on the other hand, we find
\be
\nu_{\pm 1}=\begin{cases}
	\frac{\ii}{\ellc} \,, & (\V)\,,\\
	\pm \frac{1}{\ells}+\frac{\ii}{\ellc}\,, & (\VIIh)\,.
\end{cases}
\ee
From \eqref{tilt-matching}, one notices that models $\VIIo,\V,\VIIh$
are titled, and consequently the fluid velocity has some homogeneous
vorticity. The form of this vorticity is deduced using
\be
{\rm curl} {\cal P}_i^{(\pm 1)} = \pm \nu_{\pm 1}  {\cal P}_i^{(\pm 1)} \,,
\ee
which is consistent with \eqref{Curljm}.

Finally, since Bianchi $\IX$ have compact spatial sections, the splitting into $svt$ modes is unique~\cite{King:1991jd,Pontzen2010}. Thus, all modes of the shear map uniquely to tensor perturbations of FLRW spacetimes, and there are no vector perturbations (see also Appendix~\ref{Appzeta}).

\subsection{Scalars}

Since the shear is traceless, the correspondence with metric perturbations assumes that $\phi=0$ in~\eqref{SVT-metric}. Following the same logic as for tensor and vector modes, we now look for modes $\psi$ such that
\be\label{betaijpsi}
\hatD_{ij} \psi \equiv \beta_{(0)} q^{(0)}_{ij}\,. 
\ee
The relevant equations to compare are Eqs.~\eqref{trace-free-efe} and~\eqref{trace-free-eq}. Using~\eqref{laplace-beltrami} for $j-2=|m|=0$, we arrive at
\be
{\cal S}^{(0)} = \frac{1}{3}\left[(\nu_0)^2-K \right] \,.
\ee
From Table~\ref{svtTable} we see that all open and flat models have ${\cal S}^{(0)}=0$. Thus
\be\label{matching-psi-scalar}
\nu_0 = \begin{cases}
	0 & (\I,\VIIo)\,, \\
	\frac{\ii}{\ellc} & (\V,\VIIh)\,.
\end{cases}
\ee
For the open models, $\nu_0$ corresponds to the maximal super-curvature mode. 

We stress that one will run into difficulties when finding explicitly the harmonic which leads to \eqref{betaijpsi}.  Let us consider first the flat case, corresponding to Bianchi types $\I$ or $\VIIo$, and for which $k=\nu_0=0$. One finds immediately that $\hatD^j \hatD_{ij} \psi = 0$. Hence, the moment constraint~\eqref{1st-v-constraint} which reads
\bea\label{scalar-tilt}
P_i &=&- \hatD^j \hatD_{ij} \psi'\,\nonumber\\
&=&-\frac{2}{3}(\Delta+K) \hatD_i \psi'\,,
\eea
is satisfied. However, this also implies that if \eqref{betaijpsi} is an harmonic
with $j=2,m=0$, it cannot be deduced from the STF derivative of an
harmonic with $j=1,m=0$, as in (2.22) of \cite{Pitrou:2019ifq}. The best
construction we can find consists in defining
\be\label{1st-vec-ansatz}
\psi(\gr{x}) = \frac{\beta_{(0)}}{k^2}e^{\ii\gr{k}\cdot\gr{x}}\,,\qquad k_i = k(e^3_\VIIo)_i\,.
\ee
from which we can build $\hatD_{ij} \psi$, and eventually we take the limit $k \to 0$.

For open models, the situation is slightly improved. Indeed, we can define a vector field
\be\label{2nd-vec-ansatz}
W_i \equiv -\ellc\beta_{(0)}(e^3_{\VIIh})_i\,,
\ee
such that
\be\label{2nd-vec-ansatz2}
\hatD_{\langle i} W_{j \rangle} = \beta_{(0)} q^{(0)}_{ij}\,.
\ee
One can check by direct covariant differentiation (using Eq.~\eqref{DTijkl}) that
\be\label{DiWi-derivs}
\hatD^i W_i=2\beta_{(0)}\,,\qquad \Delta W_i = -2\ellc^{-2}W_i\,.
\ee
Since $(\nu_0)^2 = -1/\ellc^2=K$ for open models, it follows from the second equation that $W_i$ is a solution of \eqref{laplace-beltrami} for $j-1=m=0$. That is, $W_i$ is an harmonic vector.

Using \eqref{2nd-vec-ansatz} in~\eqref{scalar-tilt}, we obtain
\bea
{\cal P}^{(0)}_i & = -\frac{2}{3}[(\nu_0)^2-4K]\ellc(e^3_{\VIIh})_i
\eea
where we have once again invoked~\eqref{laplace-beltrami}, this time
for $j-1=|m|=0$. We check that the above expression is satisfied for $\nu_0=\ii/\ellc$, in agreement with~\eqref{matching-psi-scalar}.
Note that the vector harmonic $W_i$ cannot be deduced from a scalar harmonic using (2.22) of \cite{Pitrou:2019ifq} as is usually the
case. Indeed, using $\nu_0 = \ii/\ellc$ (which corresponds to $k=0$) in
(A.3) of \cite{Pitrou:2019ifq}, one obtains rather immediately that the
scalar harmonics $j=0,m=0$ for that mode is a pure constant, and hence
any derivative of it vanishes. But contrary to the flat case, one needs
not consider a scalar harmonics with $k>0$ and then form the vector ($j=1,m=0$)
and tensor ($j=2,m=0$) harmonics before considering the limit $k\to 0$, since we can start our construction directly from
\eqref{2nd-vec-ansatz} using \eqref{2nd-vec-ansatz2}.

\subsection{Consistency checks}

As a final consistency check, we verify whether the curvature
perturbation and the fluid conservation equations are consistently
matched with their Bianchi counterparts in the homogeneous limit found
in the last sections. Starting with the curvature perturbation, we recall that the identification of the scalar mode assumes $\phi=0$. It then follows from \eqref{dRtroisFL} that
\begin{align}
\begin{split}\label{scalar-R3-matching}
a^2 \delta(\Rspatial) &=\frac{4}{3}\Delta (\Delta + 3 K) \psi\,,\\
&= 2 \hatD^i \hatD^j \hatD_{\langle i}W_{j \rangle}\,,
\end{split}
\end{align}
where, in going from the first to the second line, we have used the identity
\begin{align*}
\frac{2}{3}\Delta(\Delta + 3K) \psi & = \hatD^i \hatD^j \hatD_{ij} \psi\,, \\
& = \hatD^i \hatD^j \hatD_{\langle i} W_{j \rangle}\,.
\end{align*}
For flat models, $K=0=\delta(\Rspatial)$, which gives $k=\nu_0=0$, in agreement with our previous findings. For open models, the second line of~\eqref{scalar-R3-matching} can be rewritten as
\begin{align}
a^2\delta(\Rspatial) & = -\frac{4}{3}[(\nu_0)^2-4K]\hatD^iW_i\,,\nonumber \\
& = -\frac{8}{3}[(\nu_0)^2-4K]\beta_{(0)}\,,
\end{align}
where we have also used~\eqref{DiWi-derivs}. Using $\nu_0=\ii/\ellc$ from~\eqref{matching-psi-scalar} and the definition~\eqref{Rspatial-m}, we find that 
$\ellc^2{\cal R}^{(0)} = -8$, in accordance with
Table~\ref{svtTable}.

As for the fluid conservation equations, the matching follows straightforwardly. Indeed, Eqs.~\eqref{EulerEq-FL} with $\partial_i \delta p \to 0$ and~\eqref{EulerEq-Bianchi} are formally the same, whereas Eqs.~\eqref{ConservEq-FL} and~\eqref{ConservEq-Bianchi} are formally matched for $\phi=0$.

\section{Finding the $\myzeta_\ell^m$}\label{Appzeta}

We give here the technical details of the method used in computing the
constants $\myzeta_\ell^m$ needed in the matching relation \eqref{qQplus}.

\subsection{Method}\label{Methodqtilde}

Before getting into the details, we first note that relations~\eqref{OneToSevenZero} and~\eqref{FiveToSevenh} allow us to unify the description of the method using only the invariant basis of models $\I,\V,\IX$. Hence, let us define
\be\label{Defqtilde}
\tilde q^{(m)}_{ij} \equiv q^{(m)}_{kl}\tilde{M}^k_{\,\,i}\tilde{M}^l_{\,\,j}
\ee
with $\tilde{M}^i_{\,\,j} = \delta^i_j$ for models $\I$, $\V$ and $\IX$ and ${\tilde{M}^i_{\,\,j} = M^i_{\,\,j}}$ for models $\VIIo$ and $\VIIh$, where $\gr{M}$ was introduced in~\eqref{spiral-matrix}. In all cases, $\tilde q^{(0)}_{ij} = q^{(0)}_{ij}$. In this appendix, indices $i,j,\dots$ belong to the invariant co-basis of models  $\I,\V,\IX$, which are in turn associated respectively to the flat, open and closed FLRW. For instance, we have
\be\label{Useqtilde}
q^{(m)}_{ij} \gr{e}^i_{\VIIh} \otimes \gr{e}^i_{\VIIh} = \tilde q^{(m)}_{ij}\gr{e}^i_{\V} \otimes \gr{e}^j_{\V} \,.
\ee
From~\eqref{Defqtilde} we find the convenient property
\be\label{Magicdz}
\partial_z \tilde q_{ij}^{(m)} = \frac{\ii m}{\ells}\tilde q_{ij}^{(m)}\,.
\ee
Hence when computing spatial derivatives ($D_i$) of the $q_{ij}^{(m)}$
in the Bianchi $\VIIh$ case, it is convenient to use the right hand
side of \eqref{Useqtilde}. Indeed, the constants of structure associated
with the type $\V$ invariant basis have $N^{ij}=0$ and we need only
the first lines of Eqs. \eqref{UsefulDVDT1}, while the $z$-dependence
of the components $\tilde q_{ij}^{(m)}$ is handled simply via
\eqref{Magicdz}. Said differently, we avoid the complication of the
terms involving the $N^{ij}$, as their effect is equivalent to the
simple relations \eqref{Magicdz}. A similar and even simpler method applies for the
type $\VIIo$ which is related to the type $\I$ exactly as in
\eqref{Useqtilde}. Since the invariant basis associated with the $\I$ type has vanishing constants of structure,
we must only consider the derivative of the components
\eqref{Magicdz}. Hereafter we used this method extensively.

Relation \eqref{qQplus} can now be proven by proving that the relation and
all its derivatives (the covariant derivatives $\hatD_i$ associated with the background metric) hold at $\chi=0$. Checking that the relation
\eqref{qQplus} holds at $\chi=0$ is trivial as we chose the
$q_{ij}^{(m)}$ precisely on that property. Checking that the
derivatives of \eqref{qQplus} are equal at $\chi=0$ is less
obvious. However, instead of proving that all derivatives are equal at
$\chi=0$, given the general decomposition of a derivative [e.g. Eq.~(3.22) of \cite{Pitrou:2019ifq}] it is equivalent to
check only the equality of STF combinations of derivatives at the
origin, in addition to showing that the curl and the successive divergences associated with $\hatD_i$ (see \cite{Pitrou:2019ifq} for definition) of both sides of \eqref{qQplus} at $\chi=0$ are equal.

\subsection{Types $\I,\VIIo,\V,\VIIh$}

Let us consider first the $\I,\VIIo,\V,\VIIh$ cases. We first check that
the curl of an harmonic is also an harmonic [see
Ref.~\cite{Challinor:1999xz} and Eq. (3.16) of
\cite{Pitrou:2019ifq}], and similarly for plane-waves, since
\be\label{Curljm}
{\rm curl} \,{}^\ell Q^{(jm)}_{I_j}(\nu) =  \frac{m \nu}{j} \times
{}^\ell  Q^{(jm)}_{I_j}(\nu)\,,
\ee
where the curl is defined as in~\eqref{DefCurl}, but with $D_i$
replaced by $\hatD_i$. Hence the same relation is satisfied by pseudo
plane-waves. Considering the fact that the derivative
$\hatD_i$ associated with the FLRW metric is found from the spatial
derivative $D_i$ of the Bianchi metric evaluated at lowest order in
$\beta_{ij}$ (meaning in practice that we can lower and raise indices
of $A_i$ and $N^{ij}$ with $\delta_{ij}$ and $\delta^{ij}$), we can
use \eqref{CurlSFT} to obtain
\be
{\rm curl}\, q^{(m)}_{ij}=\frac{m \nu_m}{2}\times \tilde q^{(m)}_{ij}\,,
\ee
with the value \eqref{Valuenum} for $\nu_m$. Hence we get
\be
\left. {\rm curl}\, \tilde q^{(m)}_{ij}\right|_{\chi=0}=\left. \frac{\mygamma_2}{\mygamma_m}\,{\rm curl} \,Q^{(2m)}_{ij}\right|_{\chi=0}.
\ee
Next, we must also check that
\be\label{EqqQ2}
\left.\hatD_{\langle i_1} \dots \hatD_{i_n}\tilde
  q^{(m)}_{jk \rangle}\right|_{\chi=0} =\left. \frac{\mygamma_2}{\mygamma_m}\,
  \hatD_{\langle i_1} \dots \hatD_{i_n}
   Q^{(2m)}_{jk\rangle }\right|_{\chi=0}\slabel{qQplus2}\,.
 \ee
The right-hand side is evaluated using the very definition of derived
harmonics \cite{Pitrou:2019ifq} (but modified by the fact that we consider
pseudo plane-waves)
\be\label{DDQij}
\hatD_{\langle i_1} \dots \hatD_{i_{\ell-2}}  Q^{(2m)}_{i_{\ell-1} i_\ell \rangle } =
k^{\ell-2}\frac{\myzeta_\ell^m}{\myzeta_2^m} Q^{(\ell m)}_{I_\ell}\,,
\ee
together with the normalization at origin. More specifically, at
$\chi=0$, only the term $\ell=j$ in the sum \eqref{DefQfromsuml} contributes, and we then use either Eq. (2.39) or Eq. (B.30) of
\cite{Pitrou:2019ifq}. We then find for the r.h.s. of \eqref{EqqQ2}
\be
(\ii k)^{\ell-2}\frac{\mygamma_2 \zeta_\ell^m}{\mygamma_\ell
  \zeta_2^m}\left.e^z_{\langle i_1} \dots e^z_{i_{\ell-2}}\tilde
  q^{(m)}_{i_{\ell-1}i_\ell \rangle}\right|_{\chi=0}\,.\nonumber
\ee

The left hand side of \eqref{EqqQ2} is found by induction using the
method detailed in \S~\ref{Methodqtilde}, and we find
\bea\label{DDqij}
&&\left.\hatD_{\langle i_1} \dots \hatD_{i_{\ell-2}}\tilde q^{(m)}_{i_{\ell-1}i_\ell \rangle }\right|_{\chi=0}=\\
&&\prod_{p=3}^{\ell}\left(\frac{p-1}{\ellc}+\frac{m \ii}{\ells}\right)\left.e^z_{\langle i_1} \dots e^z_{i_{\ell-2}}\tilde q^{(m)}_{i_{\ell-1}i_\ell \rangle}\right|_{\chi=0}\,.\nonumber
\eea
Comparing this with \eqref{DDQij} allows to check that for $m\neq 0$, \eqref{EqqQ2} holds if
\be\label{Findzeta}
\myzeta_\ell^m = \zeta_{\ell-1}^m (-\ii)\frac{(\ell-1)+ m  \ii/\sqrt{h}}{\sqrt{\ell^2+ (\nu_m \ellc)^2}}\,.
\ee
Using the identity
\bea\label{l2nu2}
h[\ell^2 + (\nu_m \ellc)^2] &=& [(\ell-1)\sqrt{h}+m \ii]\nonumber\\
& \times& [(\ell+1)\sqrt{h} - m \ii] 
\eea
it is recast in the condition \eqref{myzetab}. However
\eqref{Findzeta} is only proven for $\ell-1 \geq 2$ with this method. To
show it is valid for $\ell-1=1$ one must check that divergences are
equal. For pseudo plane-waves the divergence is 
\be
\nabla^{p}Q^{(jm)}_{I_{j-1} p}  = -q^{(jm)} \,\frac{\myzeta_{j-1}^m}{\myzeta_j^m}\, Q^{(j-1,m)}_{I_{j-1}}\,,
\ee
where $q^{(jm)}$ is defined in Eq. (2.26) of \cite{Pitrou:2019ifq}. A direct computation then shows that
\be
\left. \hatD^j \tilde q^{(m)}_{jk}\right|_{\chi=0} =\left. \frac{\mygamma_2}{\mygamma_m}\,\hatD^j
  Q^{(2m)}_{jk}\right|_{\chi=0}\,,
\ee
implying that \eqref{Findzeta} can also be used to determine
$\myzeta_2^m/\myzeta_1^m$. Finally, for $m=0$ the equality
\be
\left. \hatD^j \hatD^k \tilde q^{(m)}_{jk}\right|_{\chi=0}
=\left. \frac{\mygamma_2}{\mygamma_m}\,\hatD^j \hatD^k
  Q^{(2m)}_{jk}\right|_{\chi=0}\,
\ee
implies that $\myzeta_1^0/\myzeta_0^0 = - \ii \sqrt{1+(\nu_0
  \ellc)^2}/2$ and thus we find \eqref{zerocase}. Note that since
$(\nu_0 \ellc)^2= - 1$, divergences must be handled as discussed in
\S~\ref{zerocase}. The global factor freedom in the $\myzeta_\ell^m$ is fixed if we choose
\eqref{myzetaa} which satisfies manifestly the necessary condition \eqref{Constraintsnuzeta}.

\subsection{Type $\IX$}\label{App9}

We now summarize the results of~Ref.~\cite{King:1991jd} so as to determine the $\myzeta_\ell^m$ in the $\IX$ model.
Let us consider a homogeneous, symmetric and trace-free tensor $\gr{T}$ on the closed FLRW background
\be
\gr{T} = T_{ij} \gr{e}^i_{\IX}\otimes \gr{e}^j_{\IX}\,.
\ee
We can prove [using e.g.~\eqref{CurlSFT}] that it must satisfy
\beas
\hatD_i T_{jk} &=& \frac{2}{3}\epsilon^p_{\,\,i (j} {\rm curl}\, T_{k)p}\,,\\
 {\rm curl}\, T_{ij}&=&\frac{3}{\ellc} T_{ij}\,. \slabel{CurlTij}
 \eeas
The first of these equations imply that $T_{ij}$ is a Killing tensor, $\hatD_{(i}T_{jk)}=0$, and in particular $\hatD^i T_{ij}=0$. From the definition of the curl, one can also show that
\be
{\rm curl}({\rm curl}\,T_{ij}) = - \Delta T_{ij} + \frac{3}{\ellc^2}T_{ij} -\frac{3}{2}\hatD_{i\langle}\hatD^kT_{j\rangle k}\,.\nonumber
\ee
If we now combine these identities together we finally arrive at
 \be
\Delta T_{ij} = -\frac{6}{\ellc^2} T_{ij}\,.
\ee
Hence, exactly as discussed in details in \cite{King:1991jd}, we find
that the homogeneous tensors on a closed FLRW are tensor harmonics with
$k^2 = 6/\ellc^2$ and hence $\nu_m^2 = (3/\ellc)^2$. Comparison with
\eqref{Curljm} then shows that it corresponds to $m \nu_m = 6/\ellc$,
and thus with $|m|=2$ the modes needed in the matching are  
\be\label{num2}
\nu_{\pm 2} = \pm \frac{3}{\ellc}\,.
\ee
Hence, all modes $q_{ij}^{(m)}$ which define the Bianchi $\IX$ as a perturbation to a closed FLRW are tensor
harmonics, or linear combination of possibly rotated tensor modes,
with  $m=2$  and $\nu_2=3/\ellc$ (along with its complex conjugate which
from table \eqref{TableParity} amounts to adding the harmonic rotated by an angle $\pi$ around the $y$
axis so as to form a standing wave of given chirality). 

Had we chosen opposite signs for the Bianchi $\IX$ constants of structure, there would be a minus sign in the right hand side of
\eqref{CurlTij}, and thus an extra minus sign in \eqref{num2}. We would find that homogeneous tensors correspond instead
to (sums of rotations of) $m=-2$ harmonics with $\nu_{-2}=3/\ellc$ (with
their complex conjugate to form standing waves). One construction
is related to the other one by a global parity transformation, and corresponds to switching the KVF with the invariant basis \cite{King:1991jd,Pontzen2010}.

\end{document}